%% file: IGRJ16318_draft.tex
\documentclass[pdftex]{pasj01}
\Received{$\langle$reception date$\rangle$}
\Accepted{$\langle$acception date$\rangle$}
\Published{$\langle$publication date$\rangle$}
\usepackage{aas_macros} 
\usepackage{graphicx} 
\newenvironment{trueauthors}{\section*{Author contributions}\fontsize{8}{11}\selectfont}{\par}

\begin{document}

\title{Glimpse of the highly obscured HMXB IGR~J16318--4848 with Hitomi
\thanks{The corresponding authors are
Hiroshi \textsc{Nakajima},
Kiyoshi \textsc{Hayashida},
Tim \textsc{Kallman},
Takuya \textsc{Miyazawa},
Hiromitsu \textsc{Takahashi},
and
Matteo \textsc{Guainazzi}
}
}
\input{hitomimember_20170821_pasj_igrj}


\email{nakajima@ess.sci.osaka-u.ac.jp}


\maketitle

\begin{abstract}

We report a Hitomi observation of IGR~J16318--4848, a high-mass X-ray binary
system with an extremely strong absorption of $N_{\rm H}$~$\sim$~$10^{24}$~cm$^{-2}$.
Previous X-ray studies revealed that its spectrum is dominated by strong
fluorescence lines of Fe as well as continuum emission.
For 
physical
and geometrical insight into the nature of the reprocessing material,
we utilize the high spectroscopic resolving power of the X-ray
microcalorimeter (the soft X-ray spectrometer; SXS) and the wide-band
sensitivity by the soft and hard X-ray imager (SXI and HXI) aboard Hitomi.
Even though photon counts are limited due to unintended off-axis
pointing, the SXS spectrum resolves Fe K$\alpha_1$ and K$\alpha_2$
lines and puts strong constraints on the line centroid and width.
The line width corresponds to the velocity of
160$^{+300}_{-70}$~km~s$^{-1}$.
This represents the most accurate, and smallest, width measurement of this line
made so far from any X-ray binary, much less than the Doppler broadening
and shift expected from speeds
which are characteristic of similar systems.
Combined with the K-shell edge energy
measured by the SXI and HXI spectra, the ionization state of Fe is
estimated to be in the range of Fe\,I--IV. 
Considering the estimated ionization parameter and the distance
between the X-ray source and the absorber,
the density and thickness of the materials are estimated.
The extraordinarily strong absorption and the absence of
a Compton shoulder component is confirmed.
These characteristics suggest reprocessing materials which
are distributed in a narrow solid angle or scattering primarily with
warm free electrons or neutral hydrogen.
This measurement was achieved using the SXS detection of 19~photons.   This provides
strong motivation for follow-up observations of this and other X-ray binaries using
the X-ray Astrophysics Recovery Mission, and other comparable future instruments.
\end{abstract}

\section{Introduction}

High-mass X-ray binaries (HMXBs) consist of a compact object
(neutron star or black hole candidate) and a massive
companion star that is typically a Be star or
a supergiant O or B type star.
HXMBs with Be companions often
show periodic variability in X-ray flux when the compact object
passes through a circumstellar decretion disk surrounding the star.
Supergiant HMXBs exhibit X-ray time variability associated with eclipse,
or partial eclipse, of the compact object by the companion star.

In addition to the comprehensive catalog of the galactic HMXBs by
\citet{2006A&A...455.1165L}, 
a recent deep survey in the hard X-ray and soft gamma-ray band
performed by IBIS/ISGRI \citep{2003A&A...411L.131U, 2003A&A...411L.141L}
onboard International
Gamma-Ray Astrophysics Laboratory (INTEGRAL) \citep{2003A&A...411L...1W}
has discovered a considerable number of HMXBs that are summarized in a catalog
by \citet{2017MNRAS.470..512K}.
More than half exhibit
persistent time variability in the hard X-ray band \citep{2013MNRAS.431..327L}.
One of the highlights of the survey is the discovery of a number of
HMXBs that exhibit extraordinarily strong absorption
with their distribution in the galaxy correlating with that of
star forming regions \citep{2012ApJ...744..108B, 2013ApJ...764..185C}.
IGR~J16318--4848 (hereafter IGR~J16318)
was the first discovered and remains the most extreme example of
such objects.

IGR~J16318 was discovered in the scanning observation of the Galactic plane by
the INTEGRAL/IBIS/ISGRI \citep{2003IAUC.8063....3C, 2003A&A...411L.427W}.
Examination of archival ASCA data revealed extremely strong X-ray
absorption toward the direction of the source \citep{2003IAUC.8070....3M}.
The X-ray spectrum is dominated by Fe K$\alpha$, K$\beta$, and Ni K$\alpha$
fluorescence emission lines and continuum
\citep{2003MNRAS.341L..13M, 2003AstL...29..644R}.
The fluorescence lines as well as the continuum vary on time scales of
thousands of seconds, corresponding to an upper limit on the emitting region
size approximately $10^{13}$~cm \citep{2003A&A...411L.427W}.

The optical/near-infrared (NIR) counterpart exhibits less absorption than
that measured in the X-ray band, which implies that the absorbing material
is concentrated around the compact object
\citep{2004ApJ...616..469F, 2005A&A...444..821L}.
The NIR spectroscopy suggests
that the counterpart is a supergiant B[e] star \citep{2004ApJ...616..469F}
based on the detection of forbidden lines of Fe.
Such stars are also known to contain dust in their envelopes
\citep{2007ApJ...667..497M};
a mid-infrared observation revealed that it is surrounded by dust and cold
gas with a heated inner rim \citep{2012ApJ...751..150C}.
The distance to the target was derived by \citet{2004ApJ...616..469F}
based on fitting of the optical/NIR spectral energy distribution (SED) fitting
to be 0.9--6.2~kpc.
\citet{2008A&A...484..801R} performed SED fitting from
optical to mid-infrared band, and utilizing the known stellar classification
of the companion star obtained a distance of 1.6~kpc.

Long term monitoring of the hard X-ray flux with Swift/BAT
shows a periodicity of $\sim$~80~d \citep{2009RAA.....9.1303J,Iyer17}.
Although the companion star belongs to the spectral type of B[e],
there is no obvious coincidence between numbers of outbursts and
orbital phase \citep{2009RAA.....9.1303J}.
Monitoring in the soft and hard X-ray band shows that the source
is always bright with flux dynamic range of a factor 5 and
Compton thick ($N_{\rm H}$~$\ge$~1.1~$\times$~$10^{24}$~cm$^{-2}$)
\citep{2010int..workE.135B}.
The statistically best spectrum obtained with Suzaku \citep{2007PASJ...59S...1M}
shows no Compton shoulder, which
implies a non-spherical and inhomogeneous absorber \citep{2009A&A...508.1275B}.
The average X-ray spectrum of the source exhibits a continuum typical
for neutron stars \citep{2004ESASP.552..417W}.
Moreover, the source shows disagreement in its X-ray/radio
flux relationship with that observed in the
low/hard state of black hole binaries
\citep{2004ApJ...616..469F}.
Nevertheless, the nature of the compact source
(neutron star or black hole candidate) is uncertain
because pulsations have not been detected.

Hitomi, the Japan-led X-ray astronomy satellite \citep{Takahashi17},
carried a microcalorimeter array (SXS; soft X-ray spectrometer)
\citep{Kelley17} which had
outstanding energy resolution in the energy band containing the Fe
K-shell lines. 
Combined with an X-ray CCD camera (SXI; soft X-ray imager) \citep{Tanaka17}
and a hard X-ray imager (HXI) \citep{Nakazawa17},
it provided unprecedented wide-band imaging spectroscopy.
Hitomi was lost due to an accident a month after the launch.
The observation of IGR~J16318 was performed during the instrument check-out phase
to demonstrate the spectroscopic performance of Hitomi.
In spite of offset pointing during the observation due to incomplete attitude
calibration, it is possible to extract significant scientific results from the
limited data.

In the remainder of this paper, we first describe the observation log including
some notes on the data
reduction in section~\ref{sec:obs}. The imaging and spectroscopic analyses
(section~\ref{sec:ana}) are followed by the discussion (section~\ref{sec:discussion})
and summary (section~\ref{sec:summary}).
Measurement errors correspond to the 90~\% confidence level, unless
otherwise indicated.

\section{Observation and data reduction}\label{sec:obs}

\subsection{Observation}\label{ssec:obs}

Pointing toward IGR~J16318 started on 22:28 10th March 2016 UT
and ended on 16:20 14th March 2016 UT.
While the SXS and SXI were already in operation, the HXI was
undergoing the startup procedure of one of the two sensors (HXI-1).
Because the observation was performed before optimizing
the alignment matrices of star trackers (STT1 and STT2),
the target was at off-axis positions throughout the observation.
The off-axis angle was \timeform{5'} according to the SXI image
after the switch of the STT from STT1 to STT2 on 17:58 13th March,
which limit the effective area of all the instruments.
The fields of view (FoV) of the SXS and HXI are \timeform{3.05'} and
\timeform{9.2'} square, respectively. Therefore only the SXI caught
the target securely within its
FoV thanks to its large FoV of \timeform{38'} square \citep{Nakajima17}.

%
The microcalorimeter array in the SXS was already in thermal
equilibrium at the time of our observation
\citep{2016SPIE.9905E..3SF, 2016SPIE.9905E..3RN}. 
The energy resolution of the onboard radioactive $^{55}$Fe source
was 4.9~eV full width half maximum (FWHM) as reported by
\citet{leutenegger17}.
However, the SXS was not in the normal operation mode in terms of some
calibration items as follows. The gate valve was still closed
and hence the effective area in the soft energy band was limited.
The Modulated X-ray Source (MXS; \cite{devries17}) was also not
yet available for contemporaneous gain measurement,
which forces us to estimate the gain uncertainty only by
onboard radioactive $^{55}$Fe sources.

The SXI was in normal operation with the "Full Window + No Burst" mode
\citep{Tanaka17}. Temperature of the CCDs was already
stable at $-110^{\circ}$C at the time of the exposure \citep{Nakajima17b}.
The observation was carried out before optimizing the parameters for
the dark level calculation and hence the SXI suffered from a cross-talk issue.
That is, an anomalously low dark level can be induced in a pixel by a charged particle
event in the adjacent segment. The dark level leads to continuous false events
in the pixel and the erroneously higher pulse heights for the normal events
around the pixel. To minimize the effect of the cross-talk issue,
the lower threshold of the effective energy band was set to be 100~ch,
which corresponds to 600~eV.

The HXI-1 completed its startup procedure and started observation on
21:30 12th March UT. The target came at the edge of the HXI-1 FoV
after the switch of the STT. Another sensor HXI-2 was still undergoing
increasing of the high voltage for the Si/CdTe double-sided
strip detectors.

\subsection{Data reduction}\label{ssec:reduction}

Hereafter, we concentrate on the data after the STT switch because
event files of all the three instruments are available in the
interval. We utilize the data cleaned and processed with a script version
03.01.005.005. All the reduction and analyses
below employ the Hitomi software version~5b and
the calibration database released on 11th May 2017
\citep{Angelini17}
\footnote{https:\slash\slash{}heasarc.gsfc.nasa.gov\slash{}docs\slash{}hitomi\slash{}calib\slash{}}.
The effective exposure times of the SXI, SXS, and HXI-1 are
39.4, 68.9, and 39.4~ks, respectively, after the data reduction.

\subsubsection{SXS}\label{sssec:sxsreduction}

\begin{figure}
 \begin{center}
  \includegraphics[width=\linewidth]{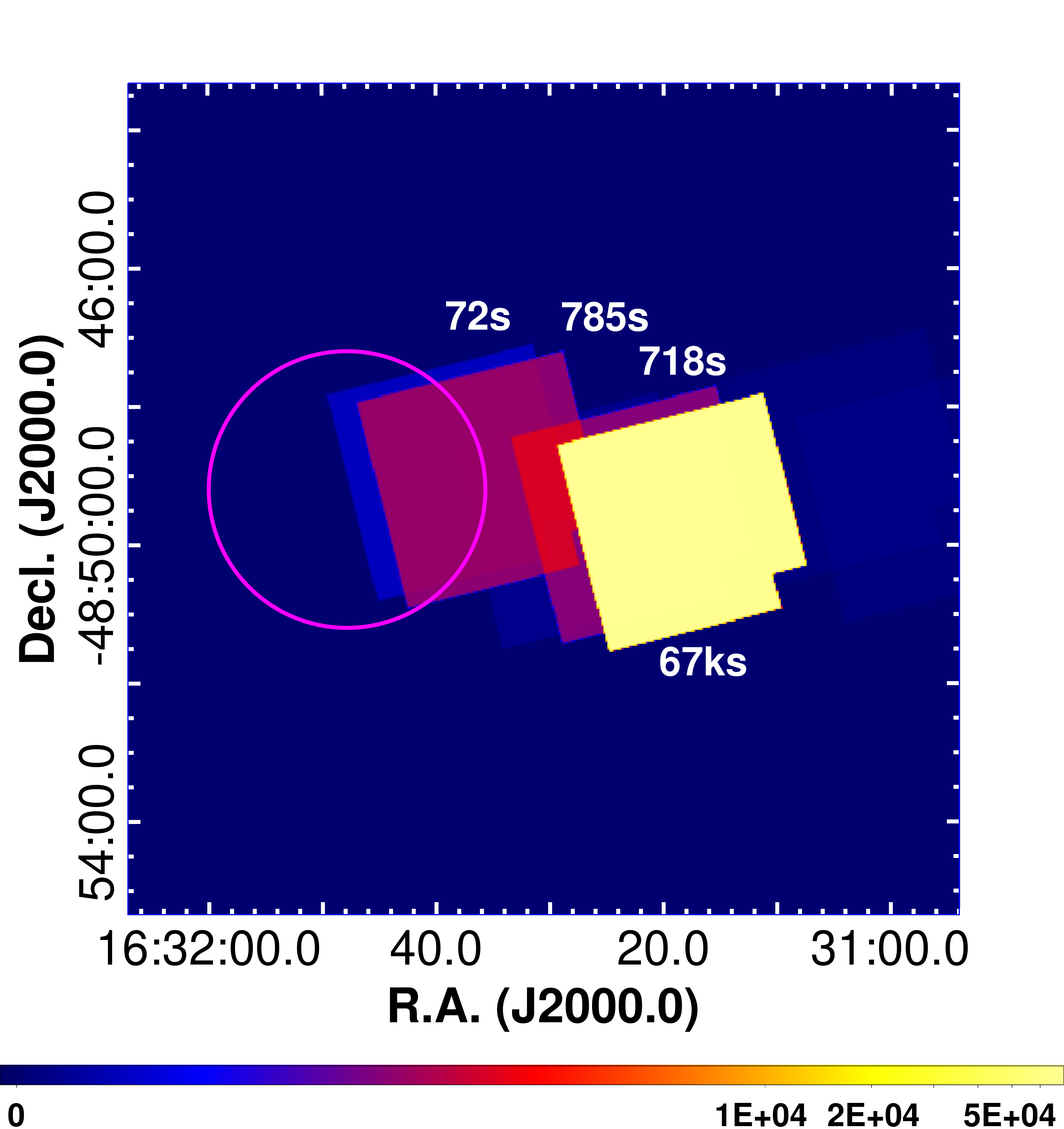}\\
  \includegraphics[width=0.75\linewidth, angle=104, origin=tr]{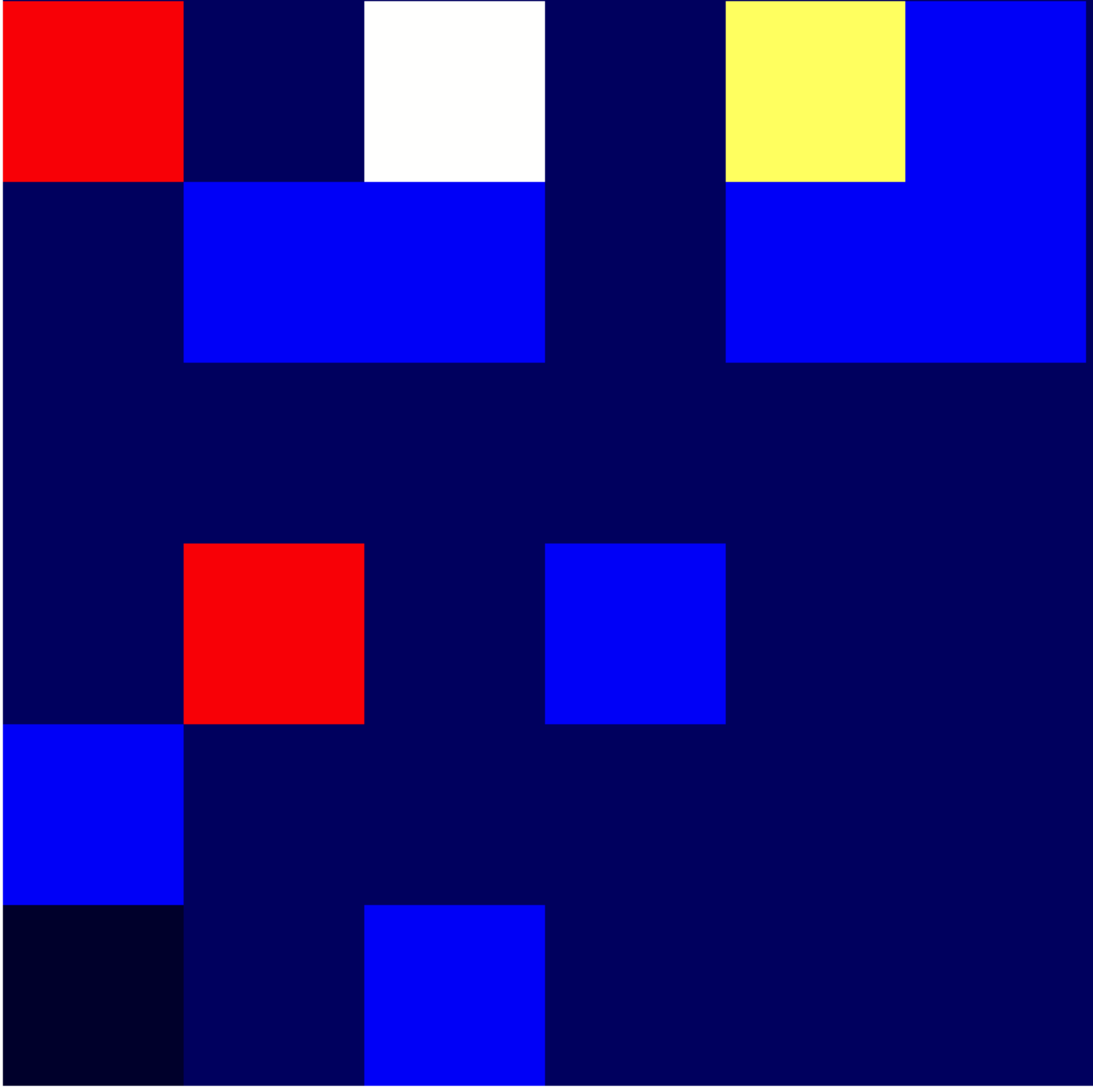}
 \end{center}
 \caption{(Top) SXS exposure map with the designation
 of the exposure time for each pointing position. The magenta circle
 corresponds to the source
 extraction region for the SXI (see figure~\ref{fig:sxiregion} bottom panel).
 (Bottom) Spatial event distribution of the SXS microcalorimeter array
 in the DET coordinate in the energy band from 6.38 to 6.42~keV.
 Blue, red, yellow and white pixels
 correspond to detection of one, two, three and four events, respectively.
 The black pixel at the bottom right is the calibration pixel
 that is not directly exposed to the sky.
 }\label{fig:sxs_expimg}
\end{figure}

Owing to the shape of the point spread function (PSF) of
the soft X-ray telescope (SXT-S; \cite{maeda17}),
some photons from the target reached the SXS in spite of the off-axis
pointing.
Furthermore, there was a wobble of the satellite at the beginning
of the observation, so that the optical axis of the SXT-S temporarily
approached the target direction. Then a part of the FoV of the SXS
overlapped with a photon extracting region for the SXI
as shown in figure~\ref{fig:sxs_expimg} top panel.

To retrieve photons from the target during the wobbling,
we relax the standard screening criteria for the angular distance between
the actual pointing and the mean pointing position (\texttt{ANG\_DIST})
from \timeform{1.5'} to \timeform{4.0'}. 
Besides the grade filtering in the standard screening,
events flagged due to close proximity in time
of 0.72~ms to other events are additionally filtered.

\begin{figure}
 \begin{center}
  \includegraphics[height=0.85\linewidth]{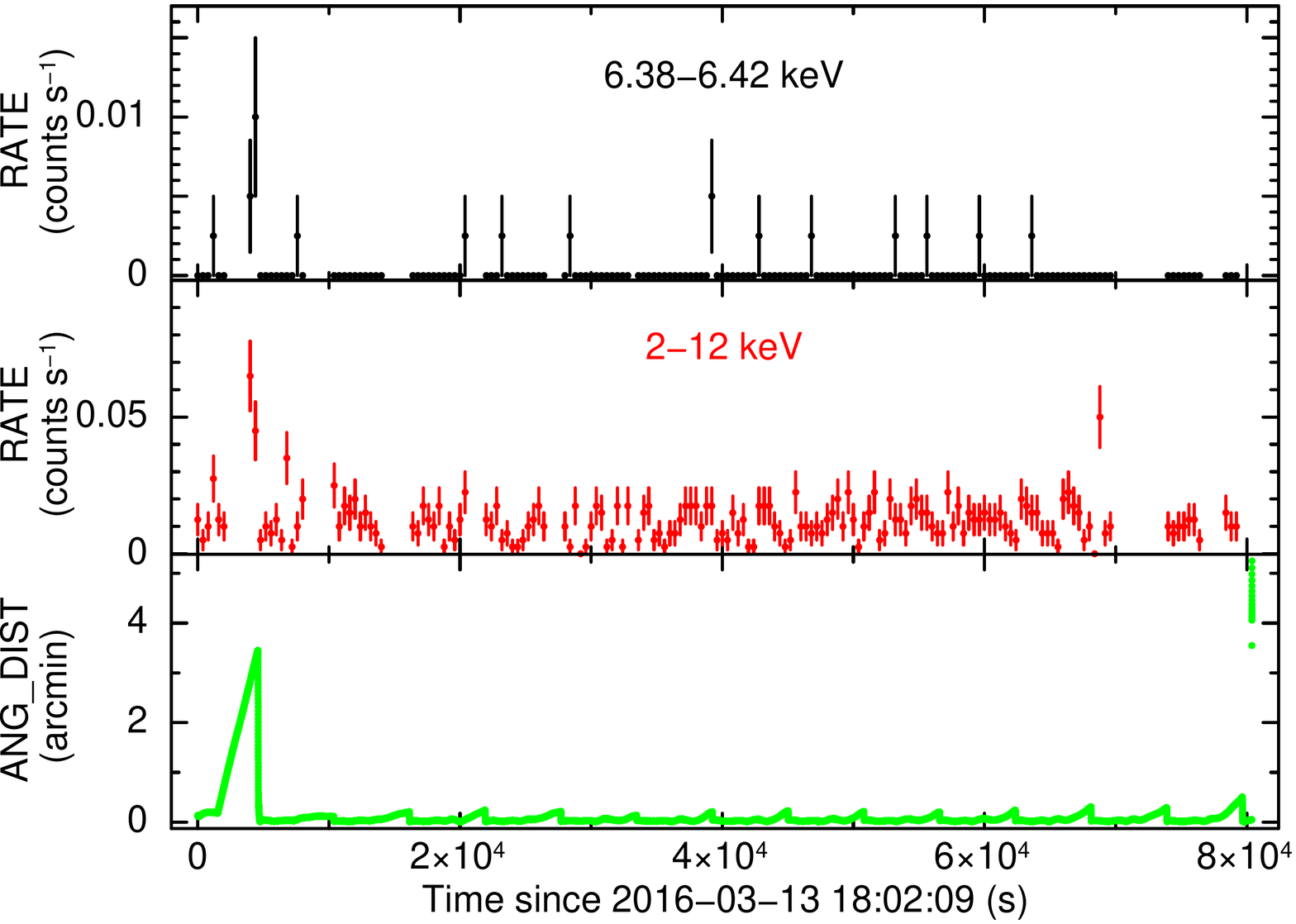}
 \end{center}
 \caption{(Top) Event light curve of the SXS full array in the energy band from
 6.38 to 6.42~keV binned with 400~s.
 (Middle) Same as the top panel but for the wide energy band
 from 2 to 12~keV. (Bottom) History of \texttt{ANG\_DIST} with
 8~s resolution.
 }\label{fig:sxs_lightcurve}
\end{figure}

Figure~\ref{fig:sxs_lightcurve} shows light curves around Fe K$\alpha$ line,
wide energy band as well as the history of the \texttt{ANG\_DIST}. The events
concentrate around the time of the wobbling in both energy bands. There
is no bright celestial target around the direction where the satellite pointed
at this time. No background flare events can be seen for other instruments
around this time. Figure~\ref{fig:sxs_expimg} bottom panel shows the spatial
distribution of the events in the energy band from 6.38 to 6.42~keV.
The 19 events spatially concentrate toward the target position.
This provides strong indication that these events originate from the target.

\subsubsection{SXI and HXI}\label{sssec:sxihxireduction}

With regard to the SXI data, false events originating from the cross talk issue
are eliminated with the parameters in \texttt{sxipipeline}
set as follows: $N_{min}$ of 6, $PHA_{sp}$ of 15, and $R$ of 0.7
\citep{Nakajima17b}.
The SXI also suffers from a light leak due to
optical/infrared light primarily when the minus Z axis of the spacecraft
points to the day earth (MZDYE). Although our observation was
free from the MZDYE periods, there was another moderate light leak
during the sun illumination of the spacecraft. We also see possible
charges left inside the CCDs after the passage of the South Atlantic
Anomaly (SAA) as described in \citet{Nakajima17b}.
The pulse heights of the events detected around the physical
edge of the CCDs are weakly affected by these issues.
The target was always near the physical edge of the CCD1
during the exposure. To minimize the effect of these problems,
we choose only the data during the eclipse of the spacecraft and
when the time after the passage of the SAA is larger than 1800~s
\citep{Nakajima17b}.
The pile-up fraction is estimated using \texttt{pileest} and the
results is below 0.7\% with a grade migration parameter of 0.1.

No additional filtering is applied to the HXI-1 cleaned event files.

\section{Analyses}\label{sec:ana}

All the spectral analyses described below are performed
using \texttt{XSPEC} v12.9.0u \citep{1996ASPC..101...17A}.
We adopt the spectral model
\texttt{tbvarabs} for the photoelectric absorption
using the interstellar medium abundances described
in \citet{2000ApJ...542..914W}.

\subsection{SXS Spectral Analysis}\label{ssec:sxs_ana}

\begin{figure}
 \begin{center}
  \includegraphics[height=0.85\linewidth]{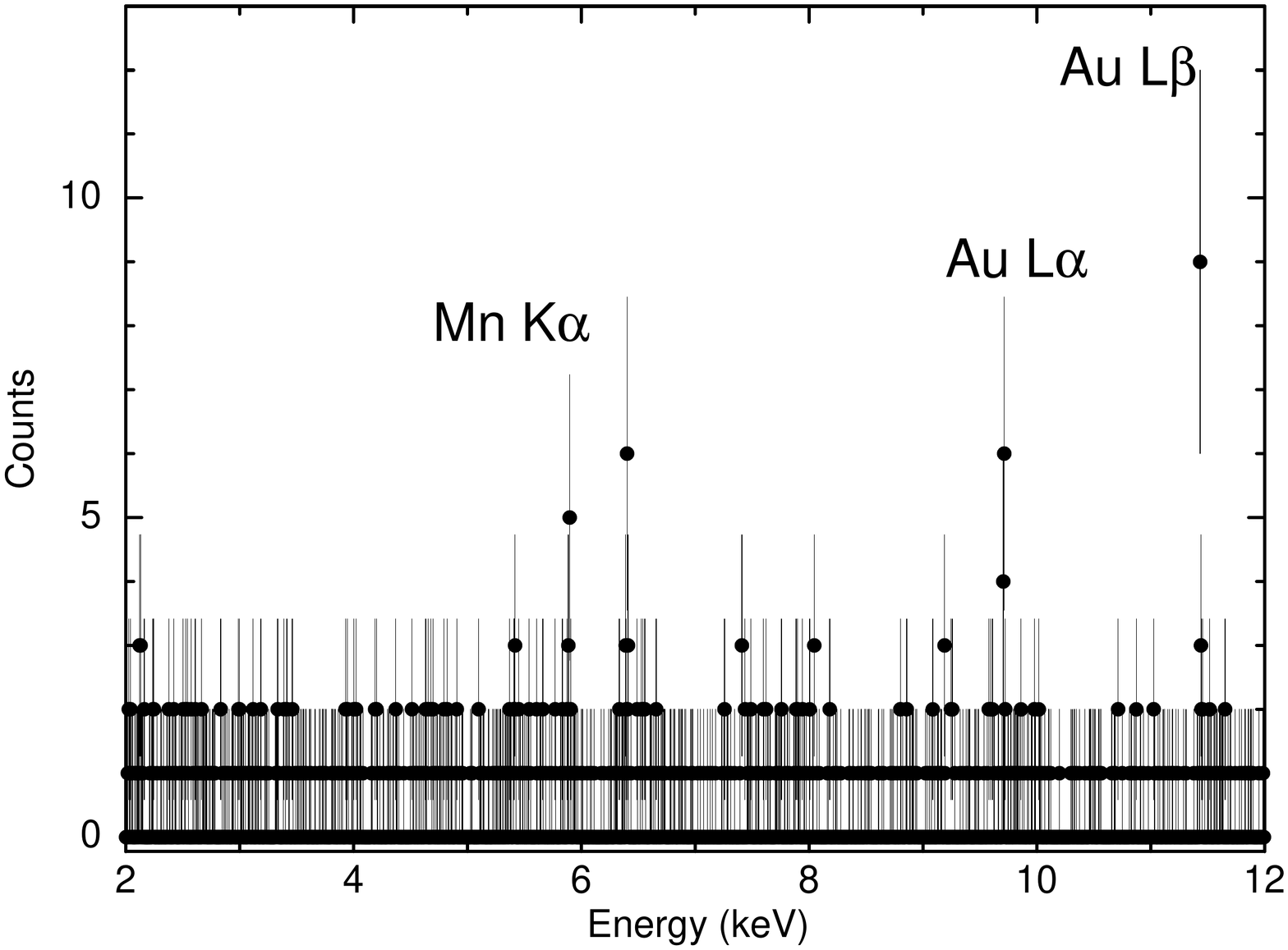}\\
  \includegraphics[height=0.85\linewidth]{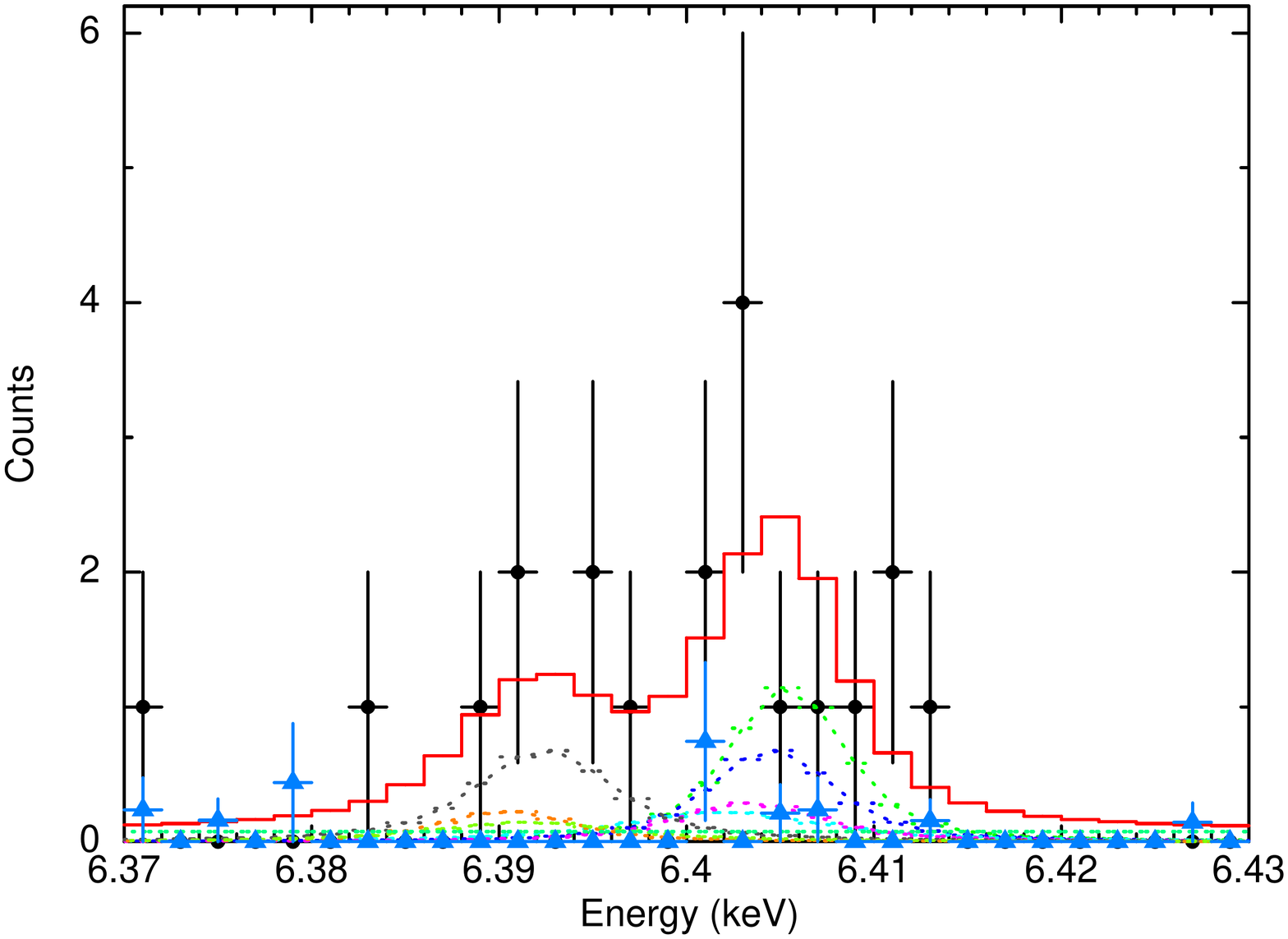}
 \end{center}
 \caption{(Top) SXS spectrum summed over all the 35 pixels.
 Peaks around 5.9, 9.7, and
 11.5~keV are the instrumental background of Mn K$\alpha$, Au L$\alpha$ and
 L$\beta$, respectively. Poisson error bars \citep{1986ApJ...303..336G}
 are presented. Note that the spectrum is binned to 4~eV.
 (Bottom) Same as the top panel but for the energy range near 6.4~keV.
 The sum of the fitted models of seven Lorentzian functions
 for the Fe K$\alpha$ lines and a power-law is shown in a solid red line,
 with each component shown in dashed lines and different colors. Although the fitting
 is performed using the original 0.5~eV per bin spectrum,
 we show the spectrum with a binning of
 2~eV for display purposes.
 Blue data with filled triangles are the calculated
 NXB spectrum that is not subtracted from the source spectrum.
 }\label{fig:sxs_spec}
\end{figure}

The spectrum obtained with the SXS in the 2--12~keV band is
shown in the top panel of figure~\ref{fig:sxs_spec}.
The events are summed over all the 35~pixels
and their total number is 752. The concentrations of events near
5.9, 9.7 and 11.5~keV originate from the instrumental background lines
of Mn K$\alpha$, Au L$\alpha$ and L$\beta$, respectively. Due to the limited
statistics of the events, we focus on the spectral analysis
around a peak at 6.4~keV that is magnified in
the bottom panel of figure~\ref{fig:sxs_spec}.
Most of the events fall within 6.39--6.41~keV
and the primary peak is slightly above 6.40~keV. This distribution
corresponds to the Fe K$\alpha_1$ and K$\alpha_2$ lines.

We estimate the number of non-X-ray background (NXB) events
\citep{kilbourne17} included in the 6.4~keV line utilizing
\texttt{sxsnxbgen}. This tool considers the magnetic cut-off
rigidity (COR) weighting of the observation and extract
events with identical filtering as the
source data from the SXS archive NXB event file.
Because the events in the energy band of 6.38--6.42~keV are
detected in the specific pixels as shown in the bottom panel
of figure~\ref{fig:sxs_expimg}, we only consider those pixels
to calculate the NXB. The estimated NXB spectrum is overlaid
on the source spectrum in the bottom panel of figure~\ref{fig:sxs_spec}.
The expected number of NXB counts in 6.38--6.42~keV
range is $\leq$~2 when we assume the same exposure time as
the target.

The K$\alpha$ line centroid near 6.4~keV implies neutral or
near-neutral ionization state of Fe. If so, the line should
be modeled with Lorentzian functions \citep{nla.cat-vn1050673}
that analytically represent the natural shape of an emission line. 
It is well known that the K$\alpha$ lines of the $3d$ transition metals
are highly asymmetric. \citet{1997PhRvA..56.4554H} applied seven
Lorentzians to accurately represent the asymmetric K$\alpha$ line
from neutral Fe.
We assume the near-neutral state and then adopt the best-fit parameters
in \citet{1997PhRvA..56.4554H},
which will be justified in section~\ref{sec:discussion}.
The NXB spectrum is represented using a power-law model with
its index fixed to zero. The power-law component is also
included to the source spectrum with its parameters linked
between the source and background.
We set the following
four parameters to be free: the energy at the maximum of the primary
Lorentzian ($\alpha_{11}$ in Table~II in \citet{1997PhRvA..56.4554H}),
its width, the normalization factor commonly multiplied to all
the seven Lorentzians, and the flux of the power-law component.
The relative energy at the maximum of each Lorentzian is fixed
as well as the relative width and amplitude. 
The continuum emissions from the target and the cosmic
X-ray background are ignored from the statistical point of view. 
We adopt c-statistics \citep{1979ApJ...228..939C} for the spectral fitting.
The original 0.5~eV per bin source and background spectra are fitted
while the binned spectra are shown in figure~\ref{fig:sxs_spec}
for display purposes.
The best-fit energy at the maximum of the primary Lorentzian is
6405.4~eV and its width is 3.5~eV (FWHM).
This yields the Fe K$\alpha_1$ line centroid of
6404.3~eV, a value which is remarkably similar with that of neutral Fe
(6403.1~eV) measured by \citet{1997PhRvA..56.4554H}.

\begin{table}
  \tbl{Best-fit parameters for the SXS spectrum.}{%
  \begin{tabular}{lc}
\hline
\hline
Parameter &  Value \\
\hline
$E_{\alpha 11}{^\ast}$ (eV)& 6405.4$\dagger$ \\
$\sigma_{\alpha 11}$ (FWHM in eV) & 3.5$\dagger$ \\
$I_{\alpha 11}$ (10$^{-4}$~cm$^{-2}$~s$^{-1}$) & 2.4 \\
$\Gamma$ & 0 (fixed) \\
$A$ (10$^{-3}$~cm$^{-2}$~s$^{-1}$)& 1.6 \\
$\rm{C}$-$\rm{stat}~(\rm{d.o.f.})$ & 131.7~(234)\\
\hline
  \end{tabular}}\label{tab:sxs_specfit}
  \begin{tabnote}
$\ast$ Energy at the maximum of the primary Lorentzian
	($\alpha_{11}$ in Table~II in \citet{1997PhRvA..56.4554H}).  \\
$\dagger$ See text for a discussion of the probability
	distributions for $E_{\alpha 11}$ and $\sigma_{\alpha 11}$.
  \end{tabnote}
\end{table}

To investigate the probability distribution function
in the parameter space, we performed Markov Chain Monte Carlo simulations within
\texttt{XSPEC}. We adopt a proposal distribution of a Gaussian for the chain with
a length of 10$^5$.
Considering the distribution, the energy at the maximum of the primary component
and its width are estimated to be 6405.4$^{+2.4}_{-2.5}$~eV
and 3.5$^{+6.4}_{-1.6}$~eV, respectively.
The best-fit parameters for the spectral fit are summarized
in table~\ref{tab:sxs_specfit}.
This is the first observational result resolving
Fe K$\alpha_1$ and K$\alpha_2$ lines for X-ray binary systems, which
demonstrates the superb energy resolution of the microcalorimeter.


The accuracy of the energy scale of the SXS is affected by
the instrumental gain uncertainty.
There had been no on-orbit full-array gain calibration before
the observation of IGR~J16318. A later calibration using
the filter-wheel $^{55}$Fe sources was carried out after
changing several cooler power settings (Eckart et al. in preparation).
Because the MXS was not yet available, a dedicated calibration pixel
that was located outside of the aperture and continuously
illuminated by a collimated $^{55}$Fe source
served as the only contemporaneous energy-scale reference.
The time-dependent scaling required to correct the gain
was applied to each pixel in the array.
It was known prior to launch that the time-dependent
gain-correction function for the calibration pixel
generally did not adequately correct the energy scale
of the array pixels. The relationship between
changes of the calibration pixel and of the array was not fixed,
but rather depended on the temperatures of the various shields
and interfaces in the dewar.
Therefore, although the relative drift rates across the array
were characterized during the later calibration with the
filter-wheel $^{55}$Fe source, the changes in cooler power
settings between the IGR~J16318 observation and that calibration
limit the usefulness of that characterization.
In fact, the measured relative gain drift predict a much larger
energy-scale offset between the final two pointings of
the Perseus cluster of galaxies than was actually observed.

To overcome our limited ability to extrapolate from the calibration
pixel, we examined the whole-array Mn K$\alpha$ instrumental line
\citep{kilbourne17} in source-free data taken from 7th March
to 15th March, when the SXS was being operated with the same cooler
settings \citep{tsujimoto17a} as those in the IGR~J16318 observation.
The SXS energy scale is found to be shifted
by at most +1~$\pm$~0.5~eV at 5.9~keV.
Further insight into the gain uncertainty comes from examining
the errors in the Mn K$\beta$ position in the filter-wheel $^{55}$Fe data after
adjusting all the pixels gain scales based on the Mn K$\alpha$ line.
The errors ranged within $-0.6$--+0.2~eV, which indicate the minimum scale
of the gain uncertainty at 6.5~keV.
We conclude that the gain shift with uncertainty of the line centroid
of Fe K$\alpha$, which is between the Mn K$\alpha$ and K$\beta$ lines,
is +1~$\pm$~0.5~eV at the time of the observation of IGR~J16318.

\subsection{SXI and HXI Analysis}\label{ssec:sxihxi_ana}

\begin{figure}
 \begin{center}
  \includegraphics[width=\linewidth]{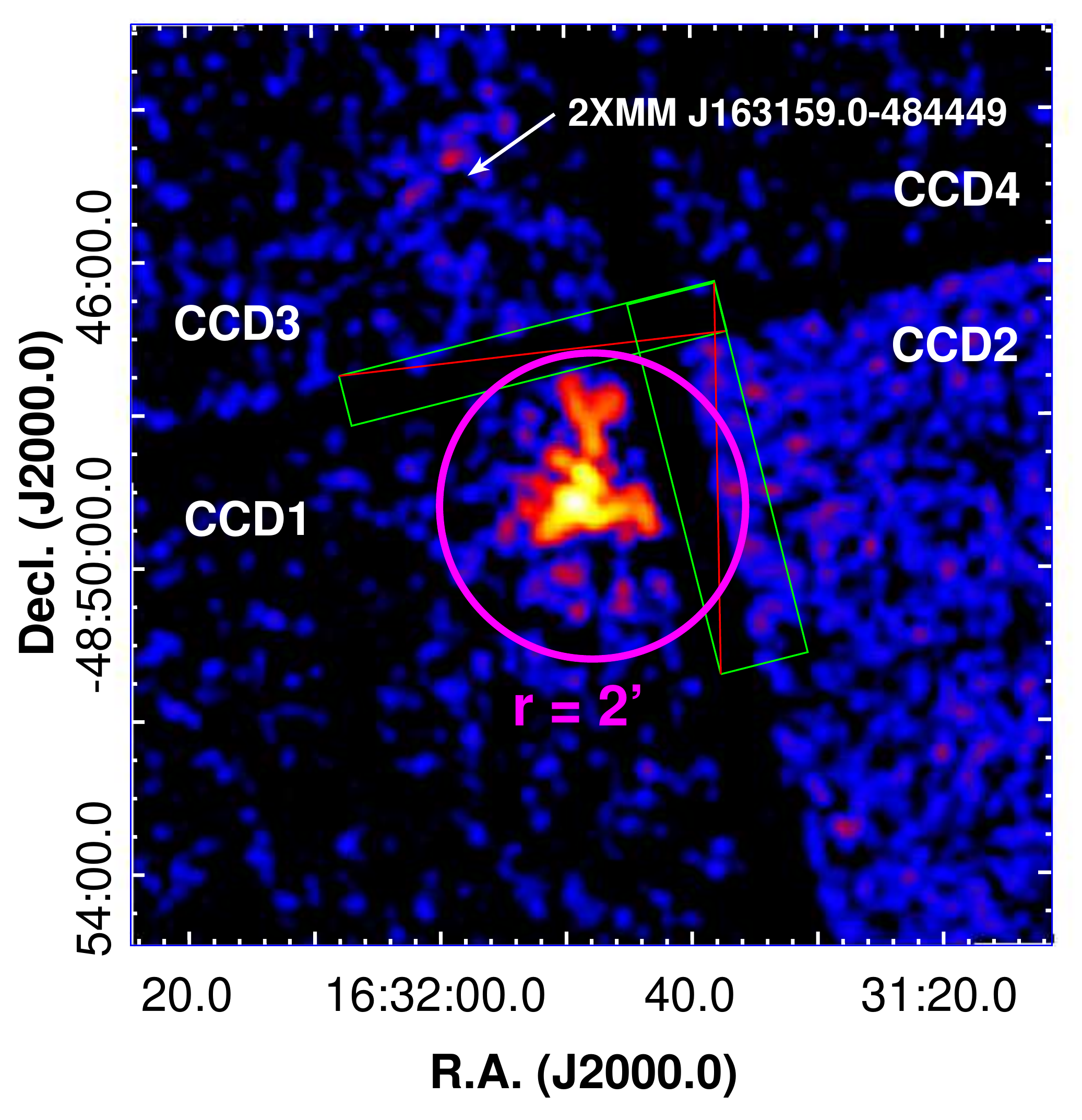}
 \end{center}
 \caption{SXI image in the energy band from 4.0 to 12~keV smoothed by
 a Gaussian of 6~pixels.
 Each CCD is designated as well as a cataloged X-ray source.
 The source spectrum extraction region is shown with a magenta circle.
 Regions shown by green rectangles with red lines are excluded
 in the extraction.
 }
 \label{fig:sxiregion}
\end{figure}

The SXI image in the energy band from 4.0 to 12~keV is shown
in figure~\ref{fig:sxiregion}.
This shows the only additional X-ray source in the FoV, based on the
2XMMi-DR3 catalog \citep{2012ApJ...756...27L}.
Note that the additional filtering of the sun illumination
of the spacecraft and the time after the passage of the SAA
is not applied to the image because the filtering has only
a small effect on the pulse height of each event.
Another note is that the PSF shape of the target is not smooth
because some pixels are affected by the cross-talk issue
\citep{Nakajima17b} and have been filtered.
In spite of the unintended off-axis pointing, the target was
securely in the CCD1.
Photon extracting regions are drawn with a magenta circle.

\begin{figure}
 \begin{center}
   \includegraphics[width=\linewidth]{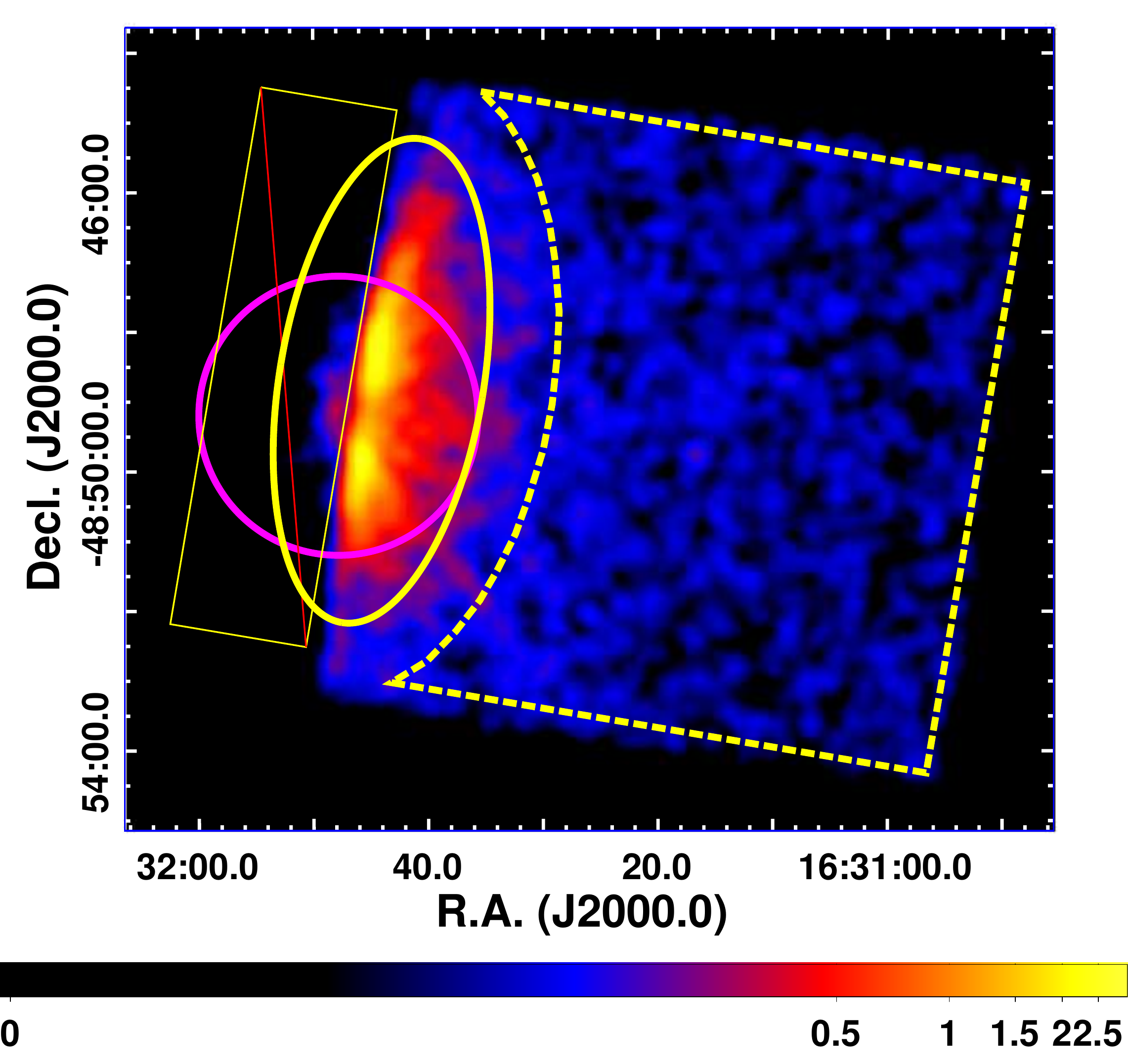}
 \end{center}
 \caption{HXI-1 image after the standard screening in the energy
 band of 5.5 to 80~keV smoothed by a Gaussian of 8~pixels.
 Source and background regions are shown with a solid ellipse
 and a dashed polygon, respectively. The same sky region as in
 figure~\ref{fig:sxiregion} is designated
 with magenta circle as a reference.
 A region shown by yellow rectangle with red line is excluded
 in the source extraction.
 }\label{fig:hxi_image}
\end{figure}

The hard X-ray image obtained by the HXI-1 in the energy band
from 5.5 to 80~keV is shown in figure~\ref{fig:hxi_image}.
The circular region in magenta designates
the same region as that in figure~\ref{fig:sxiregion}.
Thanks to the moderate PSF of the hard X-ray telescope
\citep{2016SPIE.9905E..12A}, a number of events were
detected even though the target is just on the edge of the FoV.
The source and background spectra are extracted from
the regions colored in yellow with solid and dashed lines, respectively.

\begin{figure}
 \begin{center}
   \includegraphics[height=0.85\linewidth]{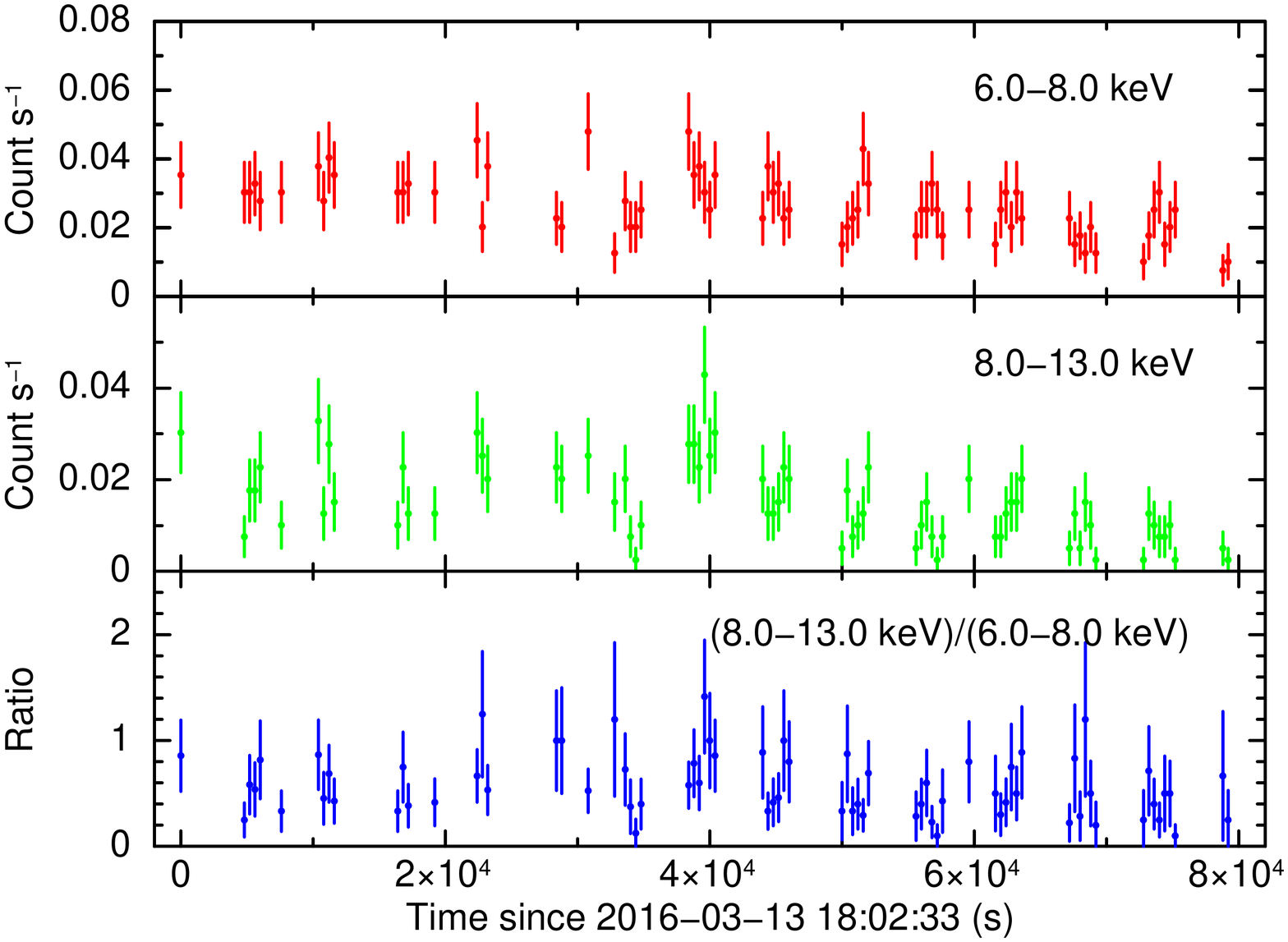}\\
   \vspace{3mm}
   \includegraphics[height=0.85\linewidth]{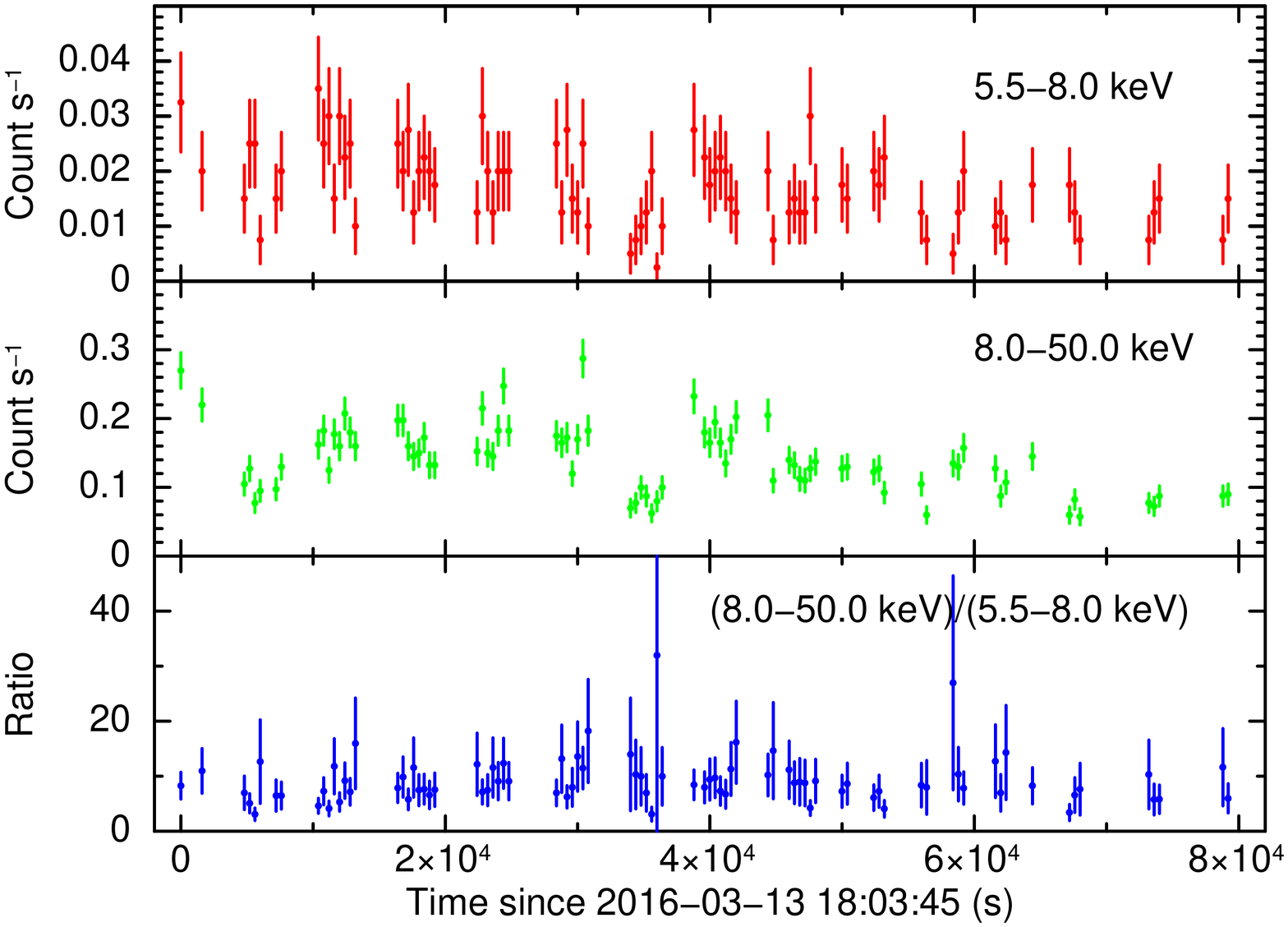}
 \end{center}
 \caption{Light curves of the SXI (top) and HXI-1 (bottom) with
 400~s resolution.
 The energy bands dominated by the fluorescence lines (red)
 and continuum emission (green)
 are shown with the ratio between the two bands (blue).
 }\label{fig:sxihxi_lightcurves}
\end{figure}

Figure~\ref{fig:sxihxi_lightcurves} shows the light curves of the SXI and
HXI-1 extracted from the source regions designated in figure~\ref{fig:sxiregion}
and figure~\ref{fig:hxi_image}, respectively.
Background is not subtracted and aspect correction is not applied.
Barycenter and dead time correction are applied for the HXI-1 data
prior to the extraction.
Note that the additional filtering of the sun illumination of the spacecraft
and the time after the passage of the SAA is not applied for the
SXI light curve because the filtering has only a small effect
on the pulse height of each event. The event rate in the energy band
dominated by
fluorescence lines and continuum both exhibit time variability on
a time scale of thousands of seconds, which is also seen
in the previous observations \citep{2007A&A...465..501I, 2009A&A...508.1275B}.
The root mean square fractional variation of the continuum band is
0.34~$\pm$~0.03 (HXI-1) and $<$~0.17 (3$\sigma$) (SXI),
while that of the fluorescence line band is $<$~0.25 (HXI-1) and
$<$~0.15 (SXI).

Pulsation search was performed both for the SXI and HXI-1 light curves in
each band shown in figure~\ref{fig:sxihxi_lightcurves} and
also in the entire band.
After the search from 1~s to one tenth of the exposure time
of each instrument, we found no significant periodic pulsation.
This prevents a conclusive determination that the compact object
is a neutron star.

\begin{figure}
 \begin{center}
   \includegraphics[height=0.85\linewidth]{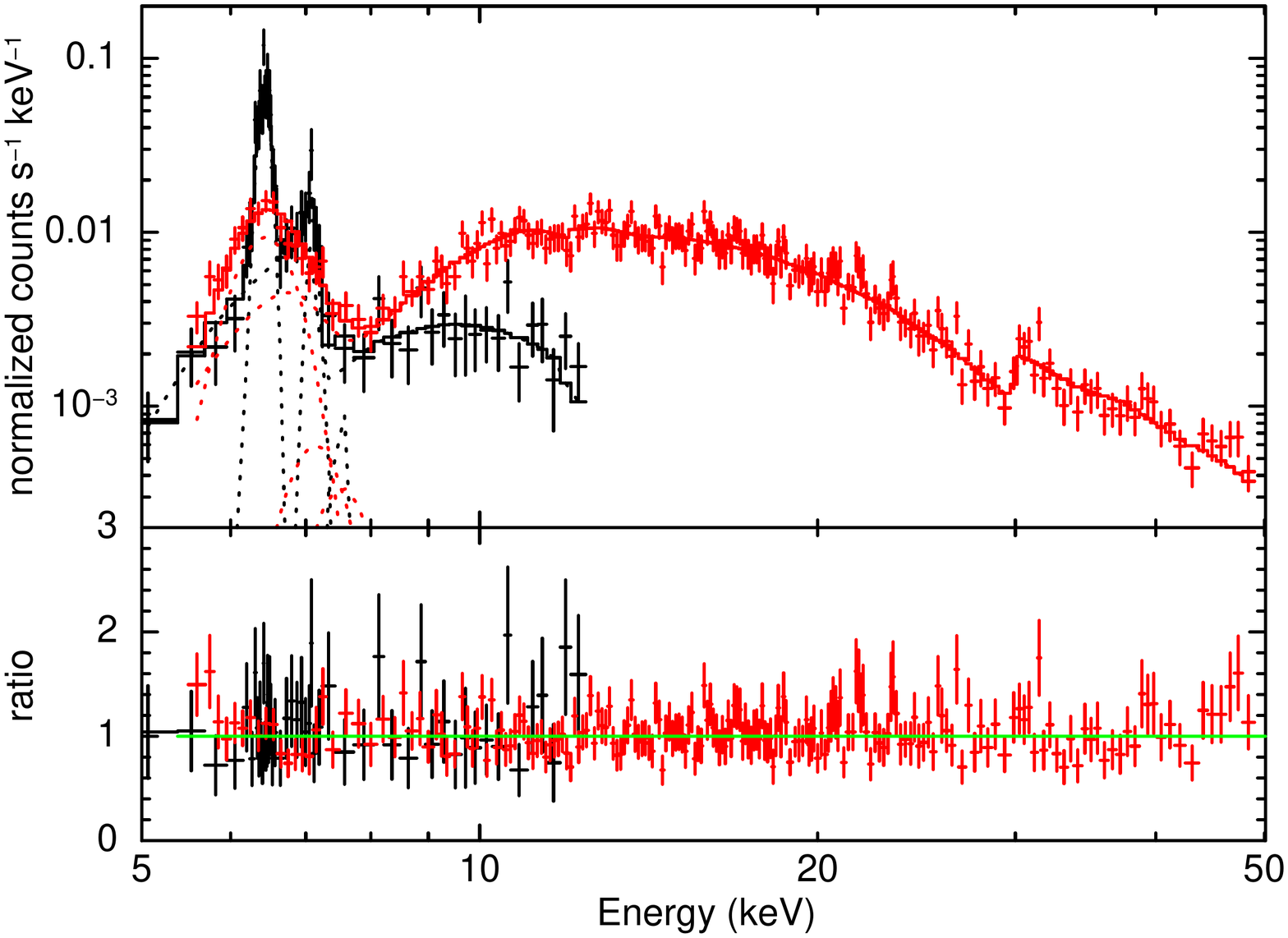}
   \includegraphics[height=0.85\linewidth]{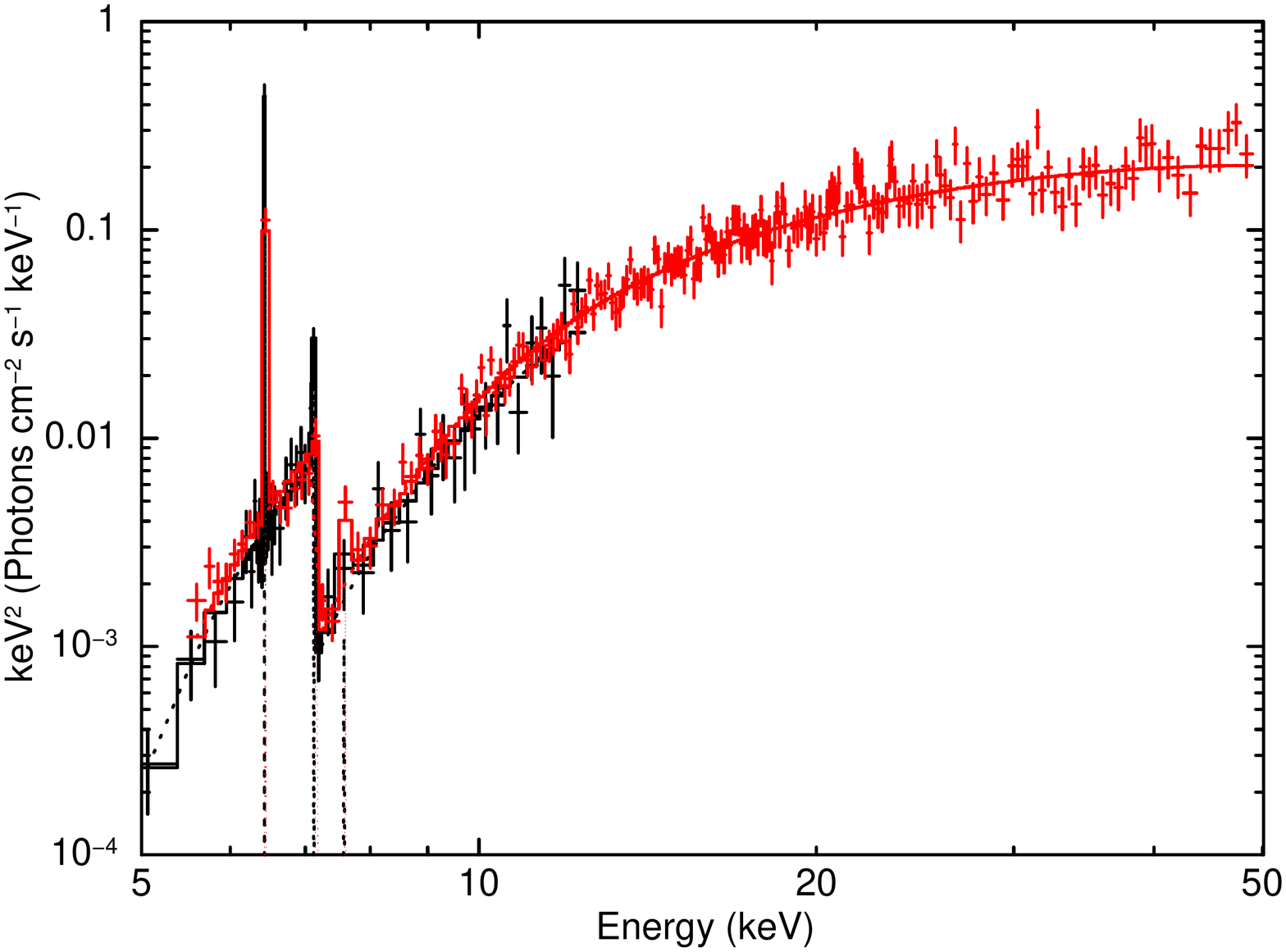} \end{center}
 \caption{(Top) Spectra obtained with the SXI (black) and HXI-1 (red).
 The best-fit spectral model is drawn with solid lines.
 Each model component is designated with dashed lines.
 (Bottom) Unfolded spectra using the best-fit
 model~A summarized in table~\ref{tab:sxihxi_specfit}. Color coding
 is the same as that in the top panel.
 }\label{fig:sxihxi_spec}
\end{figure}

Because there is no apparent outburst during the exposure, we extract the
spectra of the SXI and HXI-1 without any distinction of time.
The NXB for the SXI is calculated using \texttt{sxinxbgen} that considers
both the magnetic COR weighting of the observation and the position
of the source extracting region in the CCD.
To maximize the statistics, we subtract only the NXB component rather than
extracting background spectrum from the surrounding region for the SXI.
We extract all the events during the good time interval
of each instrument and hence the extracted durations are not precisely coincident
between the SXI and HXI-1.
In figure~\ref{fig:sxihxi_spec} top panel, we apply
a model of \texttt{tbvarabs*\{cutoffpl+gau+gau+gau\}} (hereafter model~A).
We set the Fe abundance of the absorbing material to be free
to reproduce both of the low-energy extinction and the Fe absorption edge,
while the abundances of other elements are fixed to solar values.
The difference from the model adopted in \citet{2009A&A...508.1275B}
is that we represent the fluorescence lines from the excitation states
with different total angular momenta (K$\alpha_1$ and K$\alpha_2$,
K$\beta_1$ and K$\beta_3$) with a single Gaussian function, while
\citet{2009A&A...508.1275B} introduce a Gaussian function for each
fluorescence line. Considering that the Fe K$\alpha$ line width measured with
the SXS
is negligible for the SXI and HXI-1, the widths of the Gaussian functions are
fixed to be zero. Furthermore, the line centroid of Ni K$\alpha$ is fixed
so that the ratio of the centroids of Fe K$\alpha$ and Ni K$\alpha$
becomes the value in \citet{1997PhRvA..56.4554H}.
We also introduce a constant factor that is multiplied to the HXI-1 data
to account for possible inter-instrument calibration uncertainty
of the effective area. An edge-like structure seen slightly below 30~keV
is due to an edge in quantum efficiency of the CdTe double-sided strip detectors
and hence is not seen in the unfolded spectrum shown in the bottom panel
of figure~\ref{fig:sxihxi_spec}.

The best-fit parameters are summarized in table~\ref{tab:sxihxi_specfit}.
Comparison of the spectral parameters with those
obtained from the Suzaku observation in 2006 \citep{2009A&A...508.1275B}
shows that the flux of continuum and line components significantly decreased
in the ten year interval while the equivalent widths increased.
The unabsorbed luminosity in the 2--10~keV band
is 1.0~$\times$~10$^{34}$ and 5.0~$\times$~$10^{35}$~ergs~s$^{-1}$
assuming the distance to the target of 0.9 and 6.2~kpc, respectively.
This is much less than the Eddington limit of
1.8~$\times$~$10^{38}$~ergs~s$^{-1}$
for a neutron star of 1.4~M$_{\solar}$ and is consistent with values
derived for the vast majority of HMXBs,
even if including correction for
the partial blockage of the continuum source as discussed
in section~\ref{sec:discussion}.

The Fe K-shell absorption edge energy is another key parameter that
strongly depends on the ionization state of the reprocessing materials.
In order to explore this we add the \texttt{edge} model that gives

\[
  f'(E) = \left\{ \begin{array}{ll}
    f(E) & (E < E_{\rm edge}) \\
    f(E) \cdot {\rm exp}[-\tau_{\rm MAX}(E/E_{\rm edge})^{-3}] & (E \geq E_{\rm edge}),
  \end{array} \right.
\]
where $E_{\rm edge}$ and $\tau_{\rm MAX}$ are the edge position and the absorption depth
at the edge, respectively. Because the \texttt{edge} model accounts
for absorption at the edge position, we set the Fe
abundance of the \texttt{tbvarabs} to zero in our spectral fitting.
The results are given in table~\ref{tab:sxihxi_specfit} in the column
labelled model~B.

Evaluating the flux of the possible Compton shoulder is performed by
adding another Gaussian function to model A with its centroid and width (1$\sigma$)
fixed to 6.3~keV and 50~eV, respectively \citep{2002MNRAS.337..147M}.
There is no significant flux
of the additional line with its 90\% upper limit of
5.4~$\times$~$10^{-4}$~cm$^{-2}$~s$^{-1}$ that corresponds to the 90\%
upper limit of the equivalent width of 103~eV.

\begin{table}
  \tbl{Best-fit parameters for the SXI and HXI-1 spectra.}{%
  \begin{tabular}{lcc}
\hline
\hline
Parameter &  model A & model B \\
\hline
$N_{\rm H}$ (10$^{24}$~cm$^{-2}$)& 2.06$^{+0.21}_{-0.09}$ & 2.19$^{+0.10}_{-0.06}$ \\
$A_{\rm Fe}$ & 1.19$^{+0.09}_{-0.14}$ & 0 (fixed) \\
$E_{\rm edge}$ & N/A & 7.108$^{+0.025}_{-0.046}$ \\
$\tau_{\rm MAX}$ & N/A & 2.32$^{+0.15}_{-0.26}$ \\
$\Gamma$     & 0.74$^{+0.29}_{-0.24}$ & 0.50$^{+0.02}_{-0.06}$ \\
$E_{\rm C}^\ast$  (keV) & 37.8$^{+19.3}_{-19.0}$ & 30.9$^{+10.0}_{-1.9}$ \\
$A$  (10$^{-3}$~cm$^{-2}$~s$^{-1}$) & 4.7$^{+0.3}_{-3.2}$ & 2.4$^{+0.1}_{-0.2}$  \\
$E ({\rm Fe~K\alpha})$ (keV) & 6.426$^{+0.011}_{-0.010}$ & 6.427$^{+0.011}_{-0.011}$ \\
$EW ({\rm Fe~K\alpha})$ (keV) & 2.15 & 2.09 \\
$I ({\rm Fe~K\alpha})$ (10$^{-3}$~cm$^{-2}$~s$^{-1}$) & 2.2$^{+0.8}_{-0.5}$ & 1.6$^{+0.2}_{-0.2}$ \\
$E ({\rm Fe~K\beta})$ (keV) & 7.101$^{+0.051}_{-0.001}$ & 7.108$^{+0.014}_{-0.028}$  \\
$EW ({\rm Fe~K\beta})$ (keV) & 0.38 & 0.49 \\
$I ({\rm Fe~K\beta})$ (10$^{-4}$~cm$^{-2}$~s$^{-1}$) & 1.9$^{+0.9}_{-0.7}$ & 1.8$^{+1.2}_{-0.7}$ \\
$I ({\rm Ni~K\alpha})$ (10$^{-4}$~cm$^{-2}$~s$^{-1}$) & $<$4.0 & 2.1$^{+1.8}_{-1.7}$ \\
$constant$ factor & 1.177 & 1.213 \\
$\chi^2~(\rm{d.o.f.})$ & 245.0~(251) & 250.3~(249)\\
\hline
  \end{tabular}}\label{tab:sxihxi_specfit}
  \begin{tabnote}
$\ast$ Exponential cutoff energy in the power-law model.  \\
  \end{tabnote}
\end{table}

\section{Discussion}\label{sec:discussion}

The Fe line in IGR~J16318 contains information about the ionization state
and kinematics of the emitting gas via the profile shape. It also contains
information about the quantity and geometrical distribution of the emitting
gas via the line strength, i.e., the flux or equivalent width. This does
not necessarily yield unique determinations of interesting physical quantities,
but can strongly constrain them under various scenarios. General discussions
of the dependence of flux or equivalent width have been provided by many
authors, e.g., \citet{1985gecx.conf..153K}, \citet{1986LNP...266..249M},
\citet{2010ApJ...715..947T}, and \citet{2015A&A...576A.108G}.

In particular, in the simplest case of a point source of continuum producing
the Fe K line via fluorescence at the center of a spherical uniform cloud,
simple analytic calculations show that the line equivalent width is
approximately proportional to the equivalent hydrogen column density ($N_{\rm H}$)
of the cloud for $N_{\rm H}$~$\leq$~1.5~$\times$~$10^{24}$~cm$^{-2}$.
At greater $N_{\rm H}$ the gas becomes Thomson thick and the equivalent width
no longer increases.
The maximum equivalent width is 1--2~keV and depends on the Fe elemental
abundance and on the shape of the SED of the
continuum source in the energy band above $\sim$~6~keV.
For solar Fe abundance and an SED consisting of a power-law with photon
index of 2, the maximum attainable equivalent width is less than 2~keV.
Numerical calculations for toroidal reprocessors show that the
Thomson thin approximation breaks down at
$N_{\rm H}$ much less than 1.5~$\times$~$10^{24}$~cm$^{-2}$ \citep{2010MNRAS.401..411Y}.

Equivalent widths greater than 2~keV can be obtained if the reprocessor
is not spherically symmetric around the continuum source, i.e., if 
there is an opaque screen along the direct line of sight to the
continuum source.  This is the most likely explanation for large
equivalent widths observed from X-ray binaries during eclipse
(e.g., \cite{2006ApJ...651..421W}),
or Seyfert 2 galaxies \citep{1987ApJ...320L...5K, 2016ApJ...825...85K}.
This provides a likely explanation for the large equivalent width observed from
IGR~J16318; it is crudely consistent with the column density we measure 
$N_{\rm H}$~$\simeq$~2.1~$\times$~$10^{24}$~cm$^{-2}$ together with at least
a partial blockage of the continuum source by a structure
that has Thomson depth much greater than unity.
Then we predict that the true luminosity of the source is greater than we
infer from simple dilution at a distance of 0.9--6.2~kpc, by a factor $\>$~2.

We derived the line centroid of Fe K$\alpha$ in spite of low
photon statistics. The weighted average of the energies at the maxima
of the seven Lorentzian functions is 6399.1$^{+2.5}_{-2.6}$~eV
if we consider the gain shift and uncertainty of the SXS.
Our result is consistent with those obtained
with CCD detectors aboard XMM-Newton \citep{2007A&A...465..501I}
and Suzaku \citep{2009A&A...508.1275B}. However, the uncertainty of the
measurement significantly improved with the SXS.
We have to consider the systematic velocity and the orbital velocity
of the reprocessor.
According to the NIR spectroscopy, there is no significant
systemic velocity of the companion star with
$c\Delta\lambda/\lambda$~=~$-110$~$\pm$~130~km~s$^{-1}$ \citep{2004ApJ...616..469F}.
If we assume the masses of the companion star and the compact
object of 30~${\rm M}_{\odot}$ and 1.4~${\rm M}_{\odot}$ respectively,
the line-of-sight velocity of the compact object with respect to
the companion star is within $\pm155$~km~s$^{-1}$. Then the total
Doppler velocity is expected to be $-110$~$\pm$~200~km~s$^{-1}$,
corresponding to the shift of $2.3$~$\pm4.3$~eV.

\begin{figure}
 \begin{center}
   \includegraphics[height=\linewidth]{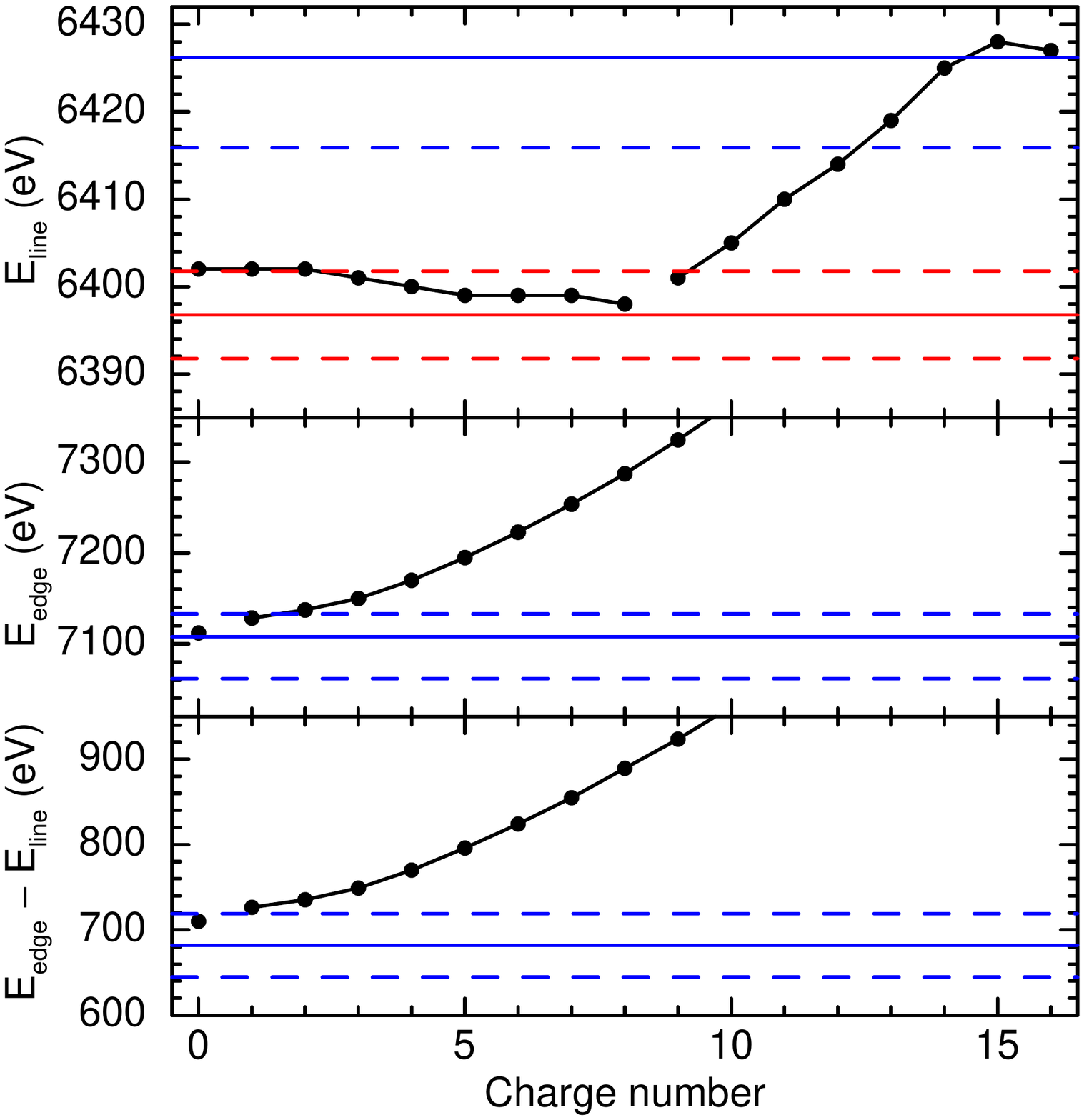}
 \end{center}
 \caption{(Top) Fe K$\alpha$ line centroid (${\rm E}_{\rm line}$)
 as a function of the ionization state
 calculated by \citet{2014ApJ...780..136Y}
 from the expectation by \citet{2003A&A...410..359P} (charge number~$\le$~8)
 and \citet{2004A&A...414..377M} (charge number~$\ge$~9).
 Values measured with the SXS and SXI+HXI-1 are shown by the red and blue
 solid lines, respectively.
 The gain shift of +1~eV and the most probable systematic velocity of the
 reprocessor are corrected for the SXS.
 The dashed lines designate 90~\% confidence level.
 (Middle) Fe K-shell ionization energy (${\rm E}_{\rm edge}$) as a function
 of the ionization state expected by \citet{2004ApJS..155..675K}
 (charge number~$\ge$~1)
 and \citet{1967RvMP...39..125B} (charge number~=~0).
 Values measured with the combined spectra of the
 SXI and HXI-1 is shown by the blue
 solid line as well as the statistical error range (dashed line).
 (Bottom) Difference of ${\rm E}_{\rm edge}$ and ${\rm E}_{\rm line}$ is plotted
 as well as the measured value with the SXI and HXI-1.
 }\label{fig:centroid_KshellE}
\end{figure}

The top panel of figure~\ref{fig:centroid_KshellE} shows the theoretical
value of the Fe K$\alpha$ line centroid (${\rm E}_{\rm line}$)
versus ionization state
\citep{2003A&A...410..359P, 2004A&A...414..377M, 2014ApJ...780..136Y}.
Comparing those with the measured values,
the ionization state of Fe\,I--X is preferred.
This is in agreement with
the other HMXBs reported by \citet{2010ApJ...715..947T}.
On the other hand, the line centroid measured with the
SXI and HXI-1 conflicts formally, at the 90\% level with that measured
with the SXS.
Monitoring the pulse heights of the onboard calibration
$^{55}$Fe source by the SXI \citep{Nakajima17b} reveals that
the pulse heights disperse in the range of $\sim$~2--3~ch
that corresponds to $\sim$~12--18~eV.
This can be interpreted as a systematic uncertainty on the SXI
energy scale and this brings the SXI+HXI-1 into marginal agreement with the SXS.

The middle panel shows the absorption edge of the Fe (${\rm E}_{\rm edge}$)
as a function of ionization state \citep{2004ApJS..155..675K, 1967RvMP...39..125B}.
The edge energy measured with the SXI and HXI-1 strongly
constrain the ionization state to be no higher than Fe\,III, which is consistent
with that obtained with the Fe K$\alpha$ line centroid.
Even if we consider the gain uncertainty of the SXI as noted above,
the ionization state is no higher than Fe\,IV.
We also plot the difference between ${\rm E}_{\rm edge}$ and ${\rm E}_{\rm line}$
in the bottom panel because such difference is rather robust against
the inaccurate energy scale.
Although the result suggests the very cold reprocessor,
Fe\,I--IV is possible if we introduce
a Doppler shift of $\sim$~1000~km~s$^{-1}$
(see later for a justification of this assumed value).
\citet{2009A&A...508.1275B} also discuss the ionization state of Fe with
the statistically best spectrum. Although their line centroid value itself
does not reject the slightly ionized state, they claim that the reprocessing
materials are neutral considering the systematic uncertainty of the gain
of Suzaku/XIS \citep{2007PASJ...59S..23K}.
Here we develop the discussion with the updated and upgraded
data obtained with Hitomi.

\citet{2004ApJS..155..675K} calculated the abundance distribution of the Fe
ions in a photoionized plasma as a function of the ionization
parameter $\xi$~=~$L/nR^2$ \citep{1969ApJ...156..943T}, where $n$
is the gas density, $R$ is the distance between the X-ray source of ionizing
radiation and the gas, and $L$ is the luminosity of the continuum emission.
The range of ionization states Fe\,I--IV is consistent with an ionization
parameter value log($\xi$)~$\lesssim$~$-2$.
The distance between X-ray source and gas responsible for the Fe emission,
$R$, can be estimated based on the X-ray time variability.
\citet{2003A&A...411L.427W} estimated the distance
to be $R$~$\simeq$~$10^{13}$~cm with XMM-Newton
by the maximum delay observed between the Fe K$\alpha$ line and
the continuum variations. Light curves obtained from other
observations \citep{2007A&A...465..501I}
also exhibited that Fe K$\alpha$ line followed
almost immediately the continuum.
Applying the $R$~$\simeq$~$10^{13}$~cm, we estimate
$n$ and the thickness of the reprocessing materials along the line
of sight ($l$) to be $n$~$\gtrsim$~3~$\times$~$10^{10}$~cm$^{-3}$ and
$l$~=~$N_{\rm H}/n$~$\lesssim$~7~$\times$~$10^{13}$~cm, respectively.
If we consider the $\sim$~80~d orbit and the masses of the companion
star and the compact object as above, the distance between them
is 2~$\times$~$10^{13}$~cm.
The maximum size of the reprocessor $l$
and $R$ may be comparable with the system size.

One of the most probable candidates for the reprocessor is the cold stellar
wind from the massive companion star.
The wind velocity ($v_w$) at the distance $r$ can be estimated
assuming the typical $\beta$-law of
\[
v_w~=~v_{\infty}(1 - R_{*}/r)^{\beta},
\]
where $v_{\infty}$ is the terminal velocity and $R_{*}$ is the
stellar radius.
Assuming the commonly used $\beta$~=~0.5 and $r$~=~$2R_{*}$,
we obtain $v_w/v_{\infty}$~$\sim$~0.7. When we assign a typical
$v_{\infty}$ of the early type stars of
$\sim$~1500--2000~km~s$^{-1}$ \citep{1978ApJ...225..893A},
$v_w$~$\sim$~1050--1400~km~s$^{-1}$ is obtained.
The measured Fe K$\alpha$ line width is equivalent to
$v$~=~160$^{+300}_{-70}$~km~s$^{-1}$.
This is much less than the Doppler
broadening expected from speeds that are characteristic of similar systems.
This indicates that the line emitting region does not cover the whole
region of the stellar wind including the companion star.
It suggests that the line
may be produced in a relatively small region centered on the compact
object. In this case, the line centroid will be Doppler-shifted
depending on the orbital phase of the compact object.
When we shift both of the line centroid and the K-shell edge energy by
25~eV that corresponds to $v_w$ of 1250~km~s$^{-1}$, the two estimates
of ionization state contradict each other.
This implies that the preferred
orbital phase is $\sim$~0.25 or $\sim$~0.75.
However, $v_{\infty}$ distributes
in a wide range even among members of the supergiant HMXBs
\citep{2016A&A...591A..26G}. Furthermore, 
\citet{2011A&A...526A..62M} argue that highly absorbed HMXB
systems have lower wind velocities than classical supergiant HMXBs.
An atmosphere model for the donor of Vela~X-1 by \citet{sander17}
also expects that the wind velocity at the neutron star location
is significantly lower than that predicted by the $\beta$-law.
More accurate determination of $v_{\infty}$
of the companion star is needed for further discussion. 
Another interesting possibility is discussed
by \citet{2015ApJ...810..102T} for supergiant HMXB. These authors argue that
Fe K$\alpha$ must be produced close to the photosphere of the donor star,
where the wind is still in the acceleration zone,
in the region facing the compact object. This case agrees with the fact that
the reprocessor does not cover the X-ray source completely.
The SXS established an empirical upper limit
to the Fe K$\alpha$ width which would imply stellar wind velocities at
distances of 1.06~$R_{*}$--1.10~$R_{*}$.
This is in agreement with theoretical predictions on the onset of wind clumps
given by \citet{2013MNRAS.428.1837S}.

\begin{figure}
 \begin{center}
   \includegraphics[height=\linewidth]{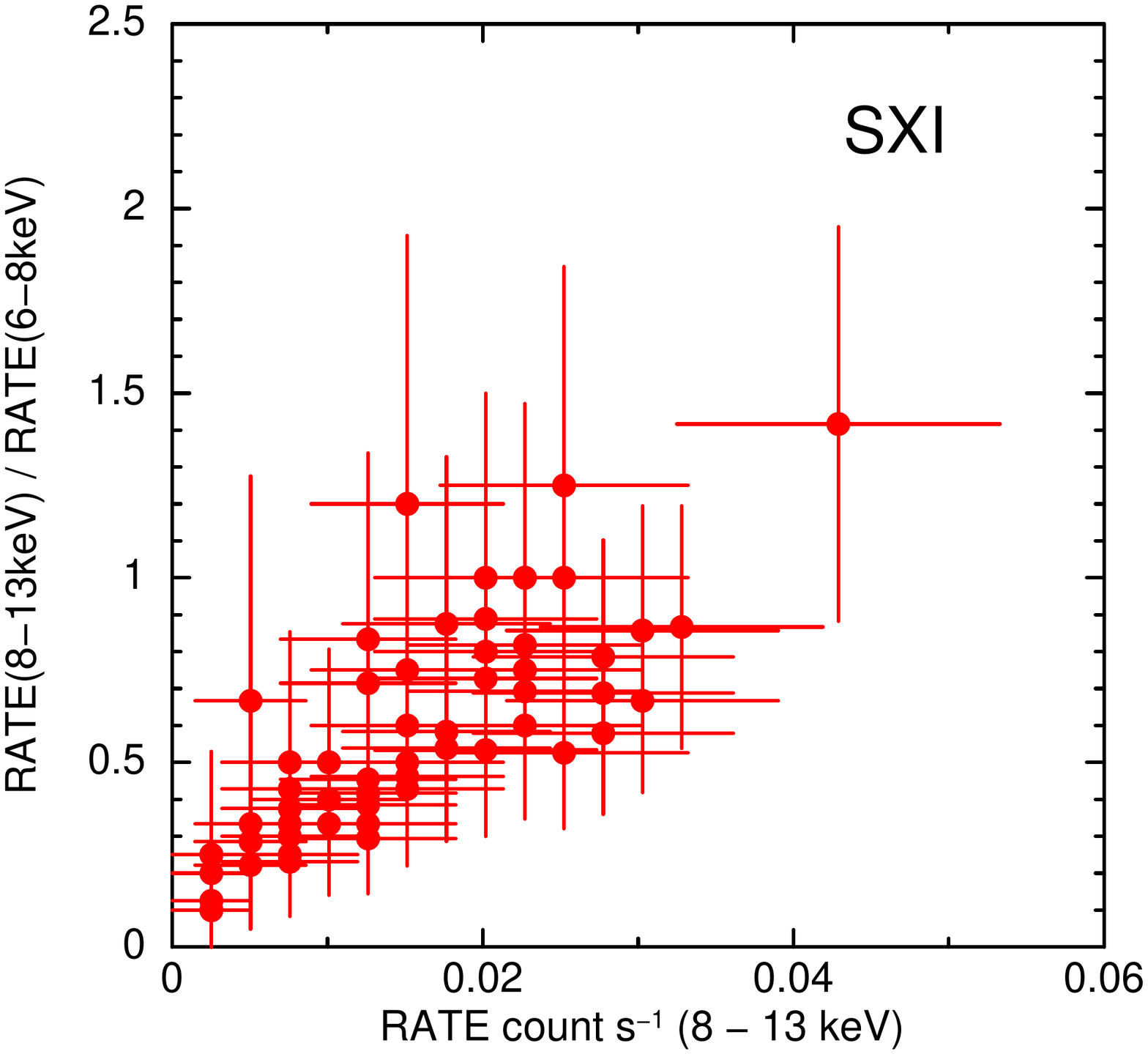}\\
   \includegraphics[height=\linewidth]{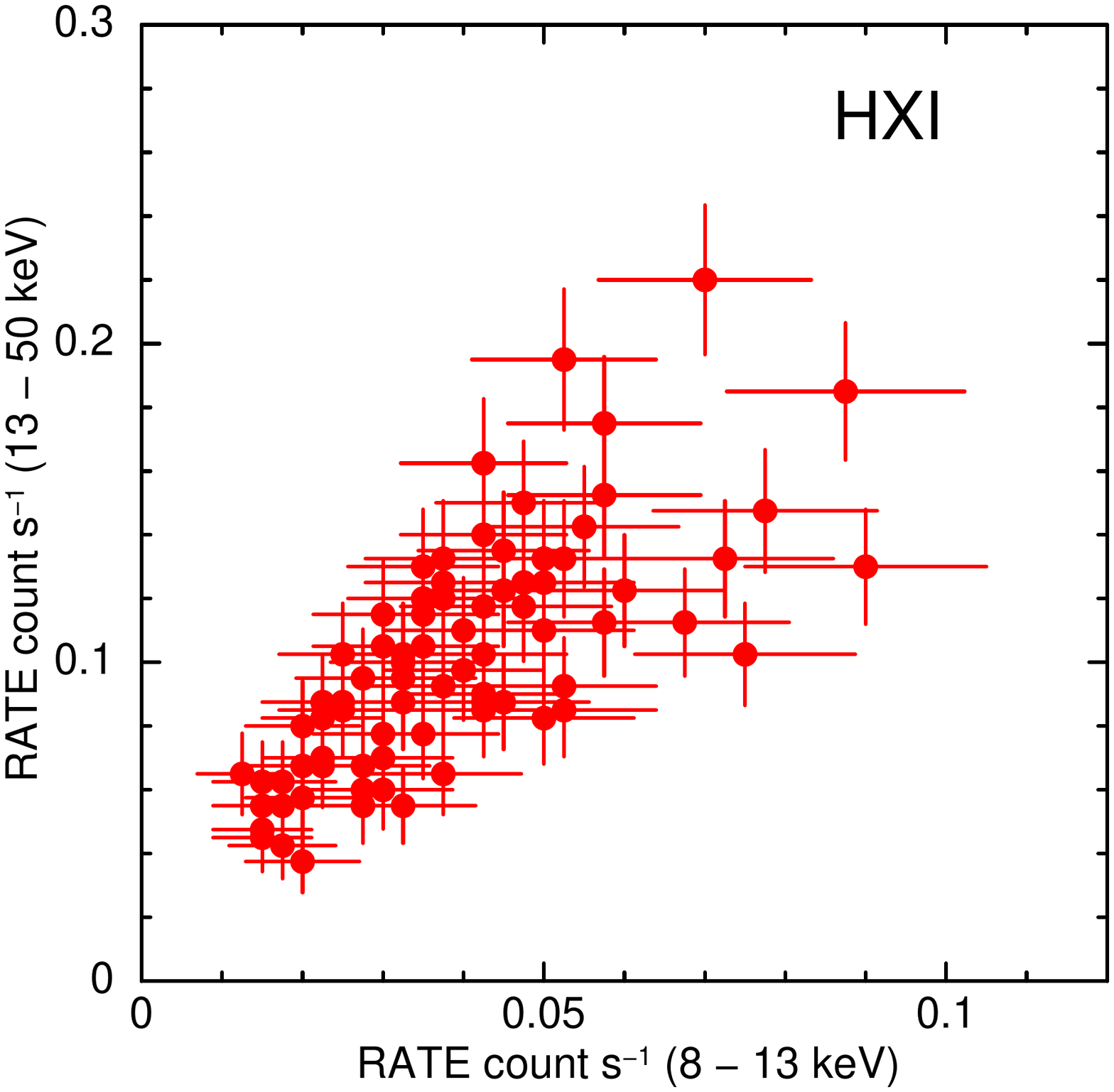}
 \end{center}
 \caption{(Top) Intensity ratio between the continuum and fluorescence line band 
 versus the intensity in the former band for the SXI. The bin size is 400~s.
 (Bottom) Count light curves of 13--50~keV band obtained with the HXI-1 versus
 that in the 8--13~keV band.
 }\label{fig:color_ratio}
\end{figure}

To investigate the time variability of the line and continuum emissions
obtained in this observation, we plot the ratio of the continuum flux
to the fluorescence line flux as a function of the latter
for the SXI in figure~\ref{fig:color_ratio} top panel.
The clear positive correlation indicates that the continuum component
exhibits variability with a larger dynamical range than
the line component, as measured with
the fractional variation of the light curves in section~\ref{ssec:sxihxi_ana}.
In other words, at least part of the line emission does not follow
the continuum variability on time scales less than 400~s.
This is consistent with the results obtained by \citet{2007A&A...465..501I}
with XMM-Newton. One possible explanation for
the positive correlation is that
the continuum is produced in a compact region while the line emission
takes place in a significantly extended region.
Another possibility is the time variation of the column density on
the line of sight.
Because the X-ray flux around the Fe K band can be affected
by the absorption column, time variation of the absorption column
on the line of sight can cause time variation only in the low energy
band. To clarify this, we check the correlation between the count
light curves in the 8--13~keV and 13--50~keV band with the HXI-1
as shown in the bottom panel of
figure~\ref{fig:color_ratio}. The clear positive correlation is a hint
of the intrinsic variation of the continuum rather than due to the changes
in the intervening column density.

The absence of the Compton shoulder is confirmed as it was in the
spectrum obtained by Suzaku \citep{2009A&A...508.1275B},
making a clear contrast with
another strongly absorbed  HMXB GX~301--2
\citep{2003ApJ...597L..37W,2011A&A...535A...9F}.
\citet{2003A&A...411L.427W} and \citet{2007A&A...465..501I} point out
that the absence of a Compton shoulder can be due to an
inhomogeneous distribution of reprocessing material.
Another possibility is the smearing of the Compton
shoulder due to the free electrons with an temperature of several eV
\citep{2003ApJ...597L..37W} and/or the scattering with neutral hydrogen
\citep{1996AstL...22..648S, 1999AstL...25..199S}.
In fact, mid-infrared observations of IGR~J16318
by \citet{2012ApJ...751..150C} revealed a spectral component
with a temperature $\sim$~37,000--40,000~K. Since this temperature
is higher than that of typical B1 supergiant stars,
they suggest that the component corresponds to
dense and hot material surrounding the stellar photosphere and
irradiated by X-rays from the compact object.
Deeper exposure with high spectral resolution like the SXS is required
for the further understanding of the circumstellar environment of this
system.

\section{Summary}\label{sec:summary}

In spite of observing challenges such as the large offset
angle and the issues such as cross-talk for the SXI, we analyze photons
from the target for all of the instruments that had been started up at
the time of the Hitomi observation of IGR~J16318.
The microcalorimeter spectrum resolved the Fe K$\alpha_1$ and K$\alpha_2$ lines
for the first time in an X-ray binary system and revealed that the
line width is narrower than that compatible with the
full range of speeds expected from a stellar wind.
Combining the line centroid measured by the SXS and the energy of the Fe K-shell
absorption edge by SXI+HXI-1, we put a constraint on the ionization state
of the reprocessing materials to be in the range of Fe\,I--IV.
Judging from the ionization parameter,
the density and thickness of the materials are estimated.
As reported in the past observations, the absorption is extraordinarily strong
($N_{\rm H}$~$>$~$10^{24}$~cm$^{-2}$) and the Compton shoulder component is not
apparent. These characteristics can be attributed to reprocessing materials
which are distributed in a narrow solid angle or scattering primarily with
warm free electrons or neutral hydrogen.

The Hitomi observation of IGR~J16318 measured the width and energy of the
Fe K fluorescence line with precision which are unprecedented for an X-ray binary.
They reveal a line width and shift which are much less than the Doppler broadening
and shift expected from speeds which are characteristic of similar systems.
This was achieved using the SXS detection of 19~photons.
If the aspect stability and accuracy of Hitomi pointing system had been
accurate at the few arc minutes level, we would have obtained far more
detailed diagnostics for the Fe K line and absorption edge diagnostics.
However this was not achieved in the initial operations of the Hitomi mission.
We now know that the physics of the Fe K line is considerably different
for this object, and perhaps for other X-ray binaries, from that previously
assumed despite over 40~years of detailed study.
Thus, microcalorimeter observations of X-ray binaries in the future with the X-ray
recovery mission will open up a new and exciting field of study.

\begin{trueauthors}
 H.~Nakajima led this research in data analysis and writing manuscript.
 He also contributed to the SXI hardware design, fabrication, integration
 and tests, in-orbit operation, and calibration.
 K.~Hayashida provided key comments on the whole discussion.
 He also made hardware and software contributions to the SXI
 as one of the instrument principal investigators.
 T.~Kallman contributed for discussion primarily on the Fe line diagnostics
 and to elaborate the manuscript.
 T.~Miyazawa worked for the fabrication and calibration of the Hard X-ray Telescope.
 H.~Takahashi contributed to the timing analyses of the HXI. He
 also made software and hardware contribution to the HXI.
 M.~Guainazzi led observation planning and gave critical comments
 mainly on the reprocessing materials.
 H.~Awaki, T.~Dotani, C.~Ferrigno, L.~C.~Gallo, P.~Gandhi, C.~A.~Kilbourne, P.~Laurent,
 K.~Mori, K.~Pottschmidt, C.~S.~Reynolds, and M.~Tsujimoto improved the manuscript.
\end{trueauthors}

\begin{ack}
We thank the support from the JSPS Core-to-Core Program.
We acknowledge all the JAXA members who have contributed to the ASTRO-H (Hitomi)
project.
All U.S. members gratefully acknowledge support through the NASA Science Mission
Directorate. Stanford and SLAC members acknowledge support via DoE contract to SLAC
National Accelerator Laboratory DE-AC3-76SF00515. Part of this work was performed under
the auspices of the U.S. DoE by LLNL under Contract DE-AC52-07NA27344.
Support from the European Space Agency is gratefully acknowledged.
French members acknowledge support from CNES, the Centre National d'\'{E}tudes Spatiales.
SRON is supported by NWO, the Netherlands Organization for Scientific Research.  Swiss
team acknowledges support of the Swiss Secretariat for Education, Research and
Innovation (SERI).
The Canadian Space Agency is acknowledged for the support of Canadian members.  
We acknowledge support from JSPS/MEXT KAKENHI grant numbers JP15J02737,
JP15H00773, JP15H00785, JP15H02070, JP15H02090, JP15H03639, JP15H03641,
JP15H03642, JP15H05438, JP15H06896, JP15K05107, JP15K17610, JP15K17657,
JP16J00548, JP16J02333, JP16H00949, JP16H03983, JP16H06342, JP16K05295,
JP16K05296, JP16K05300, JP16K05296, JP16K05309, JP16K13787, JP16K17667,
JP16K17672, JP16K17673, JP21659292, JP23340055, JP23340071, JP23540280,
JP24105007, JP24244014, JP24540232, JP24684010, JP25105516, JP25109004,
JP25247028, JP25287042, JP25400236, JP25800119, JP26109506, JP26220703,
JP26400228, JP26610047, JP26670560, and JP26800102.
The following NASA grants are acknowledged: NNX15AC76G, NNX15AE16G, NNX15AK71G,
NNX15AU54G, NNX15AW94G, and NNG15PP48P to Eureka Scientific.
H. Akamatsu acknowledges support of NWO via Veni grant.  
C. Done acknowledges STFC funding under grant ST/L00075X/1.  
A. Fabian and C. Pinto acknowledge ERC Advanced Grant 340442.
P. Gandhi acknowledges JAXA International Top Young Fellowship and UK Science and
Technology Funding Council (STFC) grant ST/J003697/2. 
Y. Ichinohe, K. Nobukawa, and H. Seta are supported by the Research Fellow of JSPS for Young
Scientists.
N. Kawai is supported by the Grant-in-Aid for Scientific Research on Innovative Areas
``New Developments in Astrophysics Through Multi-Messenger Observations of Gravitational
Wave Sources''.
S. Kitamoto is partially supported by the MEXT Supported Program for the Strategic
Research Foundation at Private Universities, 2014-2018.
B. McNamara and S. Safi-Harb acknowledge support from NSERC.
T. Dotani, T. Takahashi, T. Tamagawa, M. Tsujimoto and Y. Uchiyama acknowledge support
from the Grant-in-Aid for Scientific Research on Innovative Areas ``Nuclear Matter in
Neutron Stars Investigated by Experiments and Astronomical Observations''.
N. Werner is supported by the Lend\"ulet LP2016-11 grant from the Hungarian Academy of
Sciences.
D. Wilkins is supported by NASA through Einstein Fellowship grant number PF6-170160,
awarded by the Chandra X-ray Center, operated by the Smithsonian Astrophysical
Observatory for NASA under contract NAS8-03060.

We thank contributions by many companies, including in particular, NEC, Mitsubishi Heavy
Industries, Sumitomo Heavy Industries, and Japan Aviation Electronics Industry. Finally,
we acknowledge strong support from the following engineers.  JAXA/ISAS: Chris Baluta,
Nobutaka Bando, Atsushi Harayama, Kazuyuki Hirose, Kosei Ishimura, Naoko Iwata, Taro
Kawano, Shigeo Kawasaki, Kenji Minesugi, Chikara Natsukari, Hiroyuki Ogawa, Mina Ogawa,
Masayuki Ohta, Tsuyoshi Okazaki, Shin-ichiro Sakai, Yasuko Shibano, Maki Shida, Takanobu
Shimada, Atsushi Wada, Takahiro Yamada; JAXA/TKSC: Atsushi Okamoto, Yoichi Sato, Keisuke
Shinozaki, Hiroyuki Sugita; Chubu U: Yoshiharu Namba; Ehime U: Keiji Ogi; Kochi U of
Technology: Tatsuro Kosaka; Miyazaki U: Yusuke Nishioka; Nagoya U: Housei Nagano;
NASA/GSFC: Thomas Bialas, Kevin Boyce, Edgar Canavan, Michael DiPirro, Mark Kimball,
Candace Masters, Daniel Mcguinness, Joseph Miko, Theodore Muench, James Pontius, Peter
Shirron, Cynthia Simmons, Gary Sneiderman, Tomomi Watanabe; ADNET Systems: Michael
Witthoeft, Kristin Rutkowski, Robert S. Hill, Joseph Eggen; Wyle Information Systems:
Andrew Sargent, Michael Dutka; Noqsi Aerospace Ltd: John Doty; Stanford U/KIPAC: Makoto
Asai, Kirk Gilmore; ESA (Netherlands): Chris Jewell; SRON: Daniel Haas, Martin Frericks,
Philippe Laubert, Paul Lowes; U of Geneva: Philipp Azzarello; CSA: Alex Koujelev, Franco
Moroso.
\end{ack}

\bibliography{mybibfile}
\bibliographystyle{pasjmybst}   

\end{document}

%% file: hitomimember_20170821_pasj_igrj.tex
\author{Hitomi Collaboration,
Felix \textsc{Aharonian}\altaffilmark{1},
Hiroki \textsc{Akamatsu}\altaffilmark{2},
Fumie \textsc{Akimoto}\altaffilmark{3},
Steven W. \textsc{Allen}\altaffilmark{4,5,6},
Lorella \textsc{Angelini}\altaffilmark{7},
Marc \textsc{Audard}\altaffilmark{8},
Hisamitsu \textsc{Awaki}\altaffilmark{9},
Magnus \textsc{Axelsson}\altaffilmark{10},
Aya \textsc{Bamba}\altaffilmark{11,12},
Marshall W. \textsc{Bautz}\altaffilmark{13}
% ...
%
%
}

\altaffiltext{1}{Dublin Institute for Advanced Studies, 31 Fitzwilliam Place, Dublin 2, Ireland}
\altaffiltext{2}{SRON Netherlands Institute for Space Research, Sorbonnelaan 2, 3584 CA Utrecht, The Netherlands}
\altaffiltext{3}{Institute for Space-Earth Environmental Research, Nagoya University, Furo-cho, Chikusa-ku, Nagoya, Aichi 464-8601}
\altaffiltext{4}{Kavli Institute for Particle Astrophysics and Cosmology, Stanford University, 452 Lomita Mall, Stanford, CA 94305, USA}
\altaffiltext{5}{Department of Physics, Stanford University, 382 Via Pueblo Mall, Stanford, CA 94305, USA}
\altaffiltext{6}{SLAC National Accelerator Laboratory, 2575 Sand Hill Road, Menlo Park, CA 94025, USA}
\altaffiltext{7}{NASA, Goddard Space Flight Center, 8800 Greenbelt Road, Greenbelt, MD 20771, USA}
\altaffiltext{8}{Department of Astronomy, University of Geneva, ch. d'\'Ecogia 16, CH-1290 Versoix, Switzerland}
\altaffiltext{9}{Department of Physics, Ehime University, Bunkyo-cho, Matsuyama, Ehime 790-8577}
\altaffiltext{10}{Department of Physics and Oskar Klein Center, Stockholm University, 106 91 Stockholm, Sweden}
\altaffiltext{11}{Department of Physics, The University of Tokyo, 7-3-1 Hongo, Bunkyo-ku, Tokyo 113-0033}
\altaffiltext{12}{Research Center for the Early Universe, School of Science, The University of Tokyo, 7-3-1 Hongo, Bunkyo-ku, Tokyo 113-0033}
\altaffiltext{13}{Kavli Institute for Astrophysics and Space Research, Massachusetts Institute of Technology, 77 Massachusetts Avenue, Cambridge, MA 02139, USA}
% ...


%% file: IGRJ16318_draft.bbl
\begin{thebibliography}{73}
\providecommand{\natexlab}[1]{#1}
\expandafter\ifx\csname urlstyle\endcsname\relax
  \providecommand{\doi}[1]{doi:\discretionary{}{}{}#1}\else
  \providecommand{\doi}{doi:\discretionary{}{}{}\begingroup
  \urlstyle{rm}\Url}\fi

\bibitem[{{Abbott}(1978)}]{1978ApJ...225..893A}
{Abbott}, D.~C.
\newblock  1978, \apj,  225, 893--901.
\newblock \doi{10.1086/156554}

\bibitem[{Agarwal(1979)}]{nla.cat-vn1050673}
Agarwal, B.~K. 1979.
\newblock \emph{X-ray spectroscopy : an introduction / B. K. Agarwal}.
\newblock Springer-Verlag Berlin ; New York.
\newblock ISBN 0387092684

\bibitem[{{Angelini} et~al.(2017){Angelini}, {Terada}, {Dutka}, {Eggen},
  {Harrus}, {Hill}, and {Krimm}}]{Angelini17}
{Angelini}, L., {Terada}, Y., {Dutka}, M., {Eggen}, J., {Harrus}, I., {Hill},
  R.~S., \& {Krimm}, H.
\newblock  2017, J. Ast. Inst. Sys., submitted

\bibitem[{{Arnaud}(1996)}]{1996ASPC..101...17A}
{Arnaud}, K.~A. 1996.
\newblock In {Jacoby}, G.~H., \& {Barnes}, J., editors, \emph{Astronomical Data
  Analysis Software and Systems V},  101 of \emph{Astronomical Society of the
  Pacific Conference Series}, 17

\bibitem[{{Awaki} et~al.(2016){Awaki}, {Kunieda}, {Ishida}, {Matsumoto},
  {Furuzawa}, {Haba}, {Hayashi}, {Iizuka} et~al.}]{2016SPIE.9905E..12A}
{Awaki}, H., {Kunieda}, H., {Ishida}, M., et~al. 2016.
\newblock In \emph{Space Telescopes and Instrumentation 2016: Ultraviolet to
  Gamma Ray},  9905 of \emph{\procspie}, 990512.
\newblock \doi{10.1117/12.2231258}

\bibitem[{{Barragan} et~al.(2010){Barragan}, {Wilms}, {Kreykenbohm}, {Hanke},
  {Fuerst}, {Pottschmidt}, and {Rothschild}}]{2010int..workE.135B}
{Barragan}, L., {Wilms}, J., {Kreykenbohm}, I., {Hanke}, M., {Fuerst}, F.,
  {Pottschmidt}, K., \& {Rothschild}, R.~E. 2010.
\newblock In \emph{Eighth Integral Workshop. The Restless Gamma-ray Universe
  (INTEGRAL 2010)}, 135

\bibitem[{{Barrag{\'a}n} et~al.(2009){Barrag{\'a}n}, {Wilms}, {Pottschmidt},
  {Nowak}, {Kreykenbohm}, {Walter}, and {Tomsick}}]{2009A&A...508.1275B}
{Barrag{\'a}n}, L., {Wilms}, J., {Pottschmidt}, K., {Nowak}, M.~A.,
  {Kreykenbohm}, I., {Walter}, R., \& {Tomsick}, J.~A.
\newblock  2009, \aap,  508, 1275--1278.
\newblock \doi{10.1051/0004-6361/200810811}

\bibitem[{{Bearden} and {Burr}(1967)}]{1967RvMP...39..125B}
{Bearden}, J.~A., \& {Burr}, A.~F.
\newblock  1967, Reviews of Modern Physics, ~39, 125--142.
\newblock \doi{10.1103/RevModPhys.39.125}

\bibitem[{{Bodaghee} et~al.(2012){Bodaghee}, {Tomsick}, {Rodriguez}, and
  {James}}]{2012ApJ...744..108B}
{Bodaghee}, A., {Tomsick}, J.~A., {Rodriguez}, J., \& {James}, J.~B.
\newblock  2012, \apj,  744, 108.
\newblock \doi{10.1088/0004-637X/744/2/108}

\bibitem[{{Cash}(1979)}]{1979ApJ...228..939C}
{Cash}, W.
\newblock  1979, \apj,  228, 939--947.
\newblock \doi{10.1086/156922}

\bibitem[{{Chaty} and {Rahoui}(2012)}]{2012ApJ...751..150C}
{Chaty}, S., \& {Rahoui}, F.
\newblock  2012, \apj,  751, 150.
\newblock \doi{10.1088/0004-637X/751/2/150}

\bibitem[{{Coleiro} and {Chaty}(2013)}]{2013ApJ...764..185C}
{Coleiro}, A., \& {Chaty}, S.
\newblock  2013, \apj,  764, 185.
\newblock \doi{10.1088/0004-637X/764/2/185}

\bibitem[{{Courvoisier} et~al.(2003){Courvoisier}, {Walter}, {Rodriguez},
  {Bouchet}, and {Lutovinov}}]{2003IAUC.8063....3C}
{Courvoisier}, T.~J.-L., {Walter}, R., {Rodriguez}, J., {Bouchet}, L., \&
  {Lutovinov}, A.~A.
\newblock  2003, \iaucirc,  8063

\bibitem[{{de Vries} et~al.(2017){de Vries}, {Haas}, {Yamasaki}, {den Herder},
  {Kelley}, {Paltani}, {Kilbourne}, {Tsujimoto} et~al.}]{devries17}
{de Vries}, C.~P., {Haas}, D., {Yamasaki}, N.~Y., et~al.
\newblock  2017, J. Ast. Inst. Sys., submitted

\bibitem[{{Filliatre} and {Chaty}(2004)}]{2004ApJ...616..469F}
{Filliatre}, P., \& {Chaty}, S.
\newblock  2004, \apj,  616, 469--484.
\newblock \doi{10.1086/424869}

\bibitem[{{Fujimoto} et~al.(2016){Fujimoto}, {Takei}, {Mitsuda}, {Yamasaki},
  {Tsujimoto}, {Koyama}, {Ishikawa}, {Sugita} et~al.}]{2016SPIE.9905E..3SF}
{Fujimoto}, R., {Takei}, Y., {Mitsuda}, K., et~al. 2016.
\newblock In \emph{Space Telescopes and Instrumentation 2016: Ultraviolet to
  Gamma Ray},  9905 of \emph{\procspie}, 99053S.
\newblock \doi{10.1117/12.2232933}

\bibitem[{{F{\"u}rst} et~al.(2011){F{\"u}rst}, {Suchy}, {Kreykenbohm},
  {Barrag{\'a}n}, {Wilms}, {Pottschmidt}, {Caballero}, {Kretschmar}
  et~al.}]{2011A&A...535A...9F}
{F{\"u}rst}, F., {Suchy}, S., {Kreykenbohm}, I., et~al.
\newblock  2011, \aap,  535, A9.
\newblock \doi{10.1051/0004-6361/201117665}

\bibitem[{{Gehrels}(1986)}]{1986ApJ...303..336G}
{Gehrels}, N.
\newblock  1986, \apj,  303, 336--346.
\newblock \doi{10.1086/164079}

\bibitem[{{Gim{\'e}nez-Garc{\'{\i}}a} et~al.(2016){Gim{\'e}nez-Garc{\'{\i}}a},
  {Shenar}, {Torrej{\'o}n}, {Oskinova}, {Mart{\'{\i}}nez-N{\'u}{\~n}ez},
  {Hamann}, {Rodes-Roca}, {Gonz{\'a}lez-Gal{\'a}n}
  et~al.}]{2016A&A...591A..26G}
{Gim{\'e}nez-Garc{\'{\i}}a}, A., {Shenar}, T., {Torrej{\'o}n}, J.~M., et~al.
\newblock  2016, \aap,  591, A26.
\newblock \doi{10.1051/0004-6361/201527551}

\bibitem[{{Gim{\'e}nez-Garc{\'{\i}}a} et~al.(2015){Gim{\'e}nez-Garc{\'{\i}}a},
  {Torrej{\'o}n}, {Eikmann}, {Mart{\'{\i}}nez-N{\'u}{\~n}ez}, {Oskinova},
  {Rodes-Roca}, and {Bernab{\'e}u}}]{2015A&A...576A.108G}
{Gim{\'e}nez-Garc{\'{\i}}a}, A., {Torrej{\'o}n}, J.~M., {Eikmann}, W.,
  {Mart{\'{\i}}nez-N{\'u}{\~n}ez}, S., {Oskinova}, L.~M., {Rodes-Roca}, J.~J.,
  \& {Bernab{\'e}u}, G.
\newblock  2015, \aap,  576, A108.
\newblock \doi{10.1051/0004-6361/201425004}

\bibitem[{{H{\"o}lzer} et~al.(1997){H{\"o}lzer}, {Fritsch}, {Deutsch},
  {H{\"a}rtwig}, and {F{\"o}rster}}]{1997PhRvA..56.4554H}
{H{\"o}lzer}, G., {Fritsch}, M., {Deutsch}, M., {H{\"a}rtwig}, J., \&
  {F{\"o}rster}, E.
\newblock  1997, \pra, ~56, 4554--4568.
\newblock \doi{10.1103/PhysRevA.56.4554}

\bibitem[{{Ibarra} et~al.(2007){Ibarra}, {Matt}, {Guainazzi}, {Kuulkers},
  {Jim{\'e}nez-Bail{\'o}n}, {Rodriguez}, {Nicastro}, and
  {Walter}}]{2007A&A...465..501I}
{Ibarra}, A., {Matt}, G., {Guainazzi}, M., {Kuulkers}, E.,
  {Jim{\'e}nez-Bail{\'o}n}, E., {Rodriguez}, J., {Nicastro}, F., \& {Walter},
  R.
\newblock  2007, \aap,  465, 501--507.
\newblock \doi{10.1051/0004-6361:20066225}

\bibitem[{{Iyer} and {Paul}(2017)}]{Iyer17}
{Iyer}, N., \& {Paul}, B.
\newblock  2017, \mnras, in press.
\newblock \doi{10.1093/mnras/stx1575}

\bibitem[{{Jain} et~al.(2009){Jain}, {Paul}, and {Dutta}}]{2009RAA.....9.1303J}
{Jain}, C., {Paul}, B., \& {Dutta}, A.
\newblock  2009, Research in Astronomy and Astrophysics, ~9, 1303--1316.
\newblock \doi{10.1088/1674-4527/9/12/002}

\bibitem[{{Kallman} et~al.(2004){Kallman}, {Palmeri}, {Bautista}, {Mendoza},
  and {Krolik}}]{2004ApJS..155..675K}
{Kallman}, T.~R., {Palmeri}, P., {Bautista}, M.~A., {Mendoza}, C., \& {Krolik},
  J.~H.
\newblock  2004, \apjs,  155, 675--701.
\newblock \doi{10.1086/424039}

\bibitem[{{Kelley} et~al.(2017){Kelley}, {Mitsuda}, and {Akamatsu}}]{Kelley17}
{Kelley}, R., {Mitsuda}, K., \& {Akamatsu}, H.
\newblock  2017, J. Ast. Inst. Sys., submitted

\bibitem[{{Kilbourne} et~al.(2017){Kilbourne}, {Sawada}, {Tsujimoto},
  {Angellini}, {Boyce}, {Eckart}, {Fujimoto}, {Ishisaki} et~al.}]{kilbourne17}
{Kilbourne}, C.~A., {Sawada}, M., {Tsujimoto}, M., et~al.
\newblock  2017, \pasj

\bibitem[{{Koss} et~al.(2016){Koss}, {Assef}, {Balokovi{\'c}}, {Stern},
  {Gandhi}, {Lamperti}, {Alexander}, {Ballantyne} et~al.}]{2016ApJ...825...85K}
{Koss}, M.~J., {Assef}, R., {Balokovi{\'c}}, M., et~al.
\newblock  2016, \apj,  825, 85.
\newblock \doi{10.3847/0004-637X/825/2/85}

\bibitem[{{Koyama}(1985)}]{1985gecx.conf..153K}
{Koyama}, K. 1985.
\newblock In {Tabaka}, Y., \& {Lewin}, W.~H.~G., editors, \emph{Galactic and
  Extra-Galactic Compact X-ray Sources}, 153

\bibitem[{{Koyama} et~al.(2007){Koyama}, {Tsunemi}, {Dotani}, {Bautz},
  {Hayashida}, {Tsuru}, {Matsumoto}, {Ogawara} et~al.}]{2007PASJ...59S..23K}
{Koyama}, K., {Tsunemi}, H., {Dotani}, T., et~al.
\newblock  2007, \pasj, ~59, 23--33.
\newblock \doi{10.1093/pasj/59.sp1.S23}

\bibitem[{{Krivonos} et~al.(2017){Krivonos}, {Tsygankov}, {Mereminskiy},
  {Lutovinov}, {Sazonov}, and {Sunyaev}}]{2017MNRAS.470..512K}
{Krivonos}, R.~A., {Tsygankov}, S.~S., {Mereminskiy}, I.~A., {Lutovinov},
  A.~A., {Sazonov}, S.~Y., \& {Sunyaev}, R.~A.
\newblock  2017, \mnras,  470, 512--516.
\newblock \doi{10.1093/mnras/stx1276}

\bibitem[{{Krolik} and {Kallman}(1987)}]{1987ApJ...320L...5K}
{Krolik}, J.~H., \& {Kallman}, T.~R.
\newblock  1987, \apjl,  320, L5--L8.
\newblock \doi{10.1086/184966}

\bibitem[{{Lebrun} et~al.(2003){Lebrun}, {Leray}, {Lavocat}, {Cr{\'e}tolle},
  {Arqu{\`e}s}, {Blondel}, {Bonnin}, {Bou{\`e}re} et~al.}]{2003A&A...411L.141L}
{Lebrun}, F., {Leray}, J.~P., {Lavocat}, P., et~al.
\newblock  2003, \aap,  411, L141--L148.
\newblock \doi{10.1051/0004-6361:20031367}

\bibitem[{{Leutenegger} et~al.(2017){Leutenegger}, {Sato}, and
  {Boyce}}]{leutenegger17}
{Leutenegger}, A.~M., {Sato}, K., \& {Boyce}, K.~R.
\newblock  2017, \pasj, in preparation

\bibitem[{{Lin} et~al.(2012){Lin}, {Webb}, and {Barret}}]{2012ApJ...756...27L}
{Lin}, D., {Webb}, N.~A., \& {Barret}, D.
\newblock  2012, \apj,  756, 27.
\newblock \doi{10.1088/0004-637X/756/1/27}

\bibitem[{{Liu} et~al.(2006){Liu}, {van Paradijs}, and {van den
  Heuvel}}]{2006A&A...455.1165L}
{Liu}, Q.~Z., {van Paradijs}, J., \& {van den Heuvel}, E.~P.~J.
\newblock  2006, \aap,  455, 1165--1168.
\newblock \doi{10.1051/0004-6361:20064987}

\bibitem[{{Lutovinov} et~al.(2005){Lutovinov}, {Revnivtsev}, {Gilfanov},
  {Shtykovskiy}, {Molkov}, and {Sunyaev}}]{2005A&A...444..821L}
{Lutovinov}, A., {Revnivtsev}, M., {Gilfanov}, M., {Shtykovskiy}, P., {Molkov},
  S., \& {Sunyaev}, R.
\newblock  2005, \aap,  444, 821--829.
\newblock \doi{10.1051/0004-6361:20042392}

\bibitem[{{Lutovinov} et~al.(2013){Lutovinov}, {Revnivtsev}, {Tsygankov}, and
  {Krivonos}}]{2013MNRAS.431..327L}
{Lutovinov}, A.~A., {Revnivtsev}, M.~G., {Tsygankov}, S.~S., \& {Krivonos},
  R.~A.
\newblock  2013, \mnras,  431, 327--341.
\newblock \doi{10.1093/mnras/stt168}

\bibitem[{{Maeda} et~al.(2017){Maeda}, {Sato}, {Hayashi}, {Iizuka}, {Asai},
  {Furuzawa}, {Kurashima}, {Ishida} et~al.}]{maeda17}
{Maeda}, Y., {Sato}, T., {Hayashi}, T., et~al.
\newblock  2017, \pasj, submitted

\bibitem[{{Makishima}(1986)}]{1986LNP...266..249M}
{Makishima}, K. 1986.
\newblock In {Mason}, K.~O., {Watson}, M.~G., \& {White}, N.~E., editors,
  \emph{The Physics of Accretion onto Compact Objects},  266 of \emph{Lecture
  Notes in Physics, Berlin Springer Verlag}, 249.
\newblock \doi{10.1007/3-540-17195-9_14}

\bibitem[{{Manousakis} and {Walter}(2011)}]{2011A&A...526A..62M}
{Manousakis}, A., \& {Walter}, R.
\newblock  2011, \aap,  526, A62.
\newblock \doi{10.1051/0004-6361/201015707}

\bibitem[{{Matt}(2002)}]{2002MNRAS.337..147M}
{Matt}, G.
\newblock  2002, \mnras,  337, 147--150.
\newblock \doi{10.1046/j.1365-8711.2002.05890.x}

\bibitem[{{Matt} and {Guainazzi}(2003)}]{2003MNRAS.341L..13M}
{Matt}, G., \& {Guainazzi}, M.
\newblock  2003, \mnras,  341, L13--L17.
\newblock \doi{10.1046/j.1365-8711.2003.06658.x}

\bibitem[{{Mendoza} et~al.(2004){Mendoza}, {Kallman}, {Bautista}, and
  {Palmeri}}]{2004A&A...414..377M}
{Mendoza}, C., {Kallman}, T.~R., {Bautista}, M.~A., \& {Palmeri}, P.
\newblock  2004, \aap,  414, 377--388.
\newblock \doi{10.1051/0004-6361:20031621}

\bibitem[{{Miroshnichenko}(2007)}]{2007ApJ...667..497M}
{Miroshnichenko}, A.~S.
\newblock  2007, \apj,  667, 497--504.
\newblock \doi{10.1086/520798}

\bibitem[{{Mitsuda} et~al.(2007){Mitsuda}, {Bautz}, {Inoue}, {Kelley},
  {Koyama}, {Kunieda}, {Makishima}, {Ogawara} et~al.}]{2007PASJ...59S...1M}
{Mitsuda}, K., {Bautz}, M., {Inoue}, H., et~al.
\newblock  2007, \pasj, ~59, 1--7.
\newblock \doi{10.1093/pasj/59.sp1.S1}

\bibitem[{{Murakami} et~al.(2003){Murakami}, {Dotani}, and
  {Wijnands}}]{2003IAUC.8070....3M}
{Murakami}, H., {Dotani}, T., \& {Wijnands}, R.
\newblock  2003, \iaucirc,  8070

\bibitem[{{Nakajima}(2017)}]{Nakajima17}
{Nakajima}, H.
\newblock  2017, Nucl. Instr. and Meth. A, in press

\bibitem[{{Nakajima} et~al.(2017){Nakajima}, {Maeda}, {Uchida}, and
  {Tanaka}}]{Nakajima17b}
{Nakajima}, H., {Maeda}, Y., {Uchida}, H., \& {Tanaka}, T.
\newblock  2017, \pasj, in press

\bibitem[{{Nakazawa} et~al.(2017){Nakazawa}, {Sato}, and
  {Kokubun}}]{Nakazawa17}
{Nakazawa}, K., {Sato}, G., \& {Kokubun}, M.
\newblock  2017, J. Ast. Inst. Sys., submitted

\bibitem[{{Noda} et~al.(2016){Noda}, {Mitsuda}, {Okamoto}, {Ezoe}, {Ishikawa},
  {Fujimoto}, {Yamasaki}, {Takei} et~al.}]{2016SPIE.9905E..3RN}
{Noda}, H., {Mitsuda}, K., {Okamoto}, A., et~al. 2016.
\newblock In \emph{Space Telescopes and Instrumentation 2016: Ultraviolet to
  Gamma Ray},  9905 of \emph{\procspie}, 99053R.
\newblock \doi{10.1117/12.2231356}

\bibitem[{{Palmeri} et~al.(2003){Palmeri}, {Mendoza}, {Kallman}, {Bautista},
  and {Mel{\'e}ndez}}]{2003A&A...410..359P}
{Palmeri}, P., {Mendoza}, C., {Kallman}, T.~R., {Bautista}, M.~A., \&
  {Mel{\'e}ndez}, M.
\newblock  2003, \aap,  410, 359--364.
\newblock \doi{10.1051/0004-6361:20031262}

\bibitem[{{Rahoui} et~al.(2008){Rahoui}, {Chaty}, {Lagage}, and
  {Pantin}}]{2008A&A...484..801R}
{Rahoui}, F., {Chaty}, S., {Lagage}, P.-O., \& {Pantin}, E.
\newblock  2008, \aap,  484, 801--813.
\newblock \doi{10.1051/0004-6361:20078774}

\bibitem[{{Revnivtsev}(2003)}]{2003AstL...29..644R}
{Revnivtsev}, M.~G.
\newblock  2003, Astronomy Letters, ~29, 644--648.
\newblock \doi{10.1134/1.1615332}

\bibitem[{{Sander} et~al.(2017){Sander}, {F{\"u}rst}, {Kretschmar}, {Oskinova},
  {Todt}, {Hainich}, {Shenar}, and {Hamann}}]{sander17}
{Sander}, A.~A.~C., {F{\"u}rst}, F., {Kretschmar}, P., {Oskinova}, L.~M.,
  {Todt}, H., {Hainich}, R., {Shenar}, T., \& {Hamann}, W.-R.
\newblock  2017, \aap, submitted

\bibitem[{{Sundqvist} and {Owocki}(2013)}]{2013MNRAS.428.1837S}
{Sundqvist}, J.~O., \& {Owocki}, S.~P.
\newblock  2013, \mnras,  428, 1837--1844.
\newblock \doi{10.1093/mnras/sts165}

\bibitem[{{Sunyaev} and {Churazov}(1996)}]{1996AstL...22..648S}
{Sunyaev}, R.~A., \& {Churazov}, E.~M.
\newblock  1996, Astronomy Letters, ~22, 648--663

\bibitem[{{Sunyaev} et~al.(1999){Sunyaev}, {Uskov}, and
  {Churazov}}]{1999AstL...25..199S}
{Sunyaev}, R.~A., {Uskov}, D.~B., \& {Churazov}, E.~M.
\newblock  1999, Astronomy Letters, ~25, 199--205

\bibitem[{{Takahashi} et~al.(2017){Takahashi}, {Kokubun}, and
  {Mitsuda}}]{Takahashi17}
{Takahashi}, T., {Kokubun}, M., \& {Mitsuda}, H.
\newblock  2017, J. Ast. Inst. Sys., submitted

\bibitem[{{Tanaka} et~al.(2017){Tanaka}, {Uchida}, and {Nakajima}}]{Tanaka17}
{Tanaka}, T., {Uchida}, H., \& {Nakajima}, H.
\newblock  2017, J. Ast. Inst. Sys., submitted

\bibitem[{{Tarter} et~al.(1969){Tarter}, {Tucker}, and
  {Salpeter}}]{1969ApJ...156..943T}
{Tarter}, C.~B., {Tucker}, W.~H., \& {Salpeter}, E.~E.
\newblock  1969, \apj,  156, 943.
\newblock \doi{10.1086/150026}

\bibitem[{{Torrej{\'o}n} et~al.(2010){Torrej{\'o}n}, {Schulz}, {Nowak}, and
  {Kallman}}]{2010ApJ...715..947T}
{Torrej{\'o}n}, J.~M., {Schulz}, N.~S., {Nowak}, M.~A., \& {Kallman}, T.~R.
\newblock  2010, \apj,  715, 947--958.
\newblock \doi{10.1088/0004-637X/715/2/947}

\bibitem[{{Torrej{\'o}n} et~al.(2015){Torrej{\'o}n}, {Schulz}, {Nowak},
  {Oskinova}, {Rodes-Roca}, {Shenar}, and {Wilms}}]{2015ApJ...810..102T}
{Torrej{\'o}n}, J.~M., {Schulz}, N.~S., {Nowak}, M.~A., {Oskinova}, L.,
  {Rodes-Roca}, J.~J., {Shenar}, T., \& {Wilms}, J.
\newblock  2015, \apj,  810, 102.
\newblock \doi{10.1088/0004-637X/810/2/102}

\bibitem[{{Tsujimoto} et~al.(2017){Tsujimoto}, {Mitsuda}, {Kelley}, {den
  Herder}, and et~al.}]{tsujimoto17a}
{Tsujimoto}, M., {Mitsuda}, K., {Kelley}, R.~L., {den Herder}, J.~W., \& et~al.
\newblock  2017, J. Ast. Inst. Sys., submitted

\bibitem[{{Ubertini} et~al.(2003){Ubertini}, {Lebrun}, {Di Cocco}, {Bazzano},
  {Bird}, {Broenstad}, {Goldwurm}, {La Rosa} et~al.}]{2003A&A...411L.131U}
{Ubertini}, P., {Lebrun}, F., {Di Cocco}, G., et~al.
\newblock  2003, \aap,  411, L131--L139.
\newblock \doi{10.1051/0004-6361:20031224}

\bibitem[{{Walter} et~al.(2004){Walter}, {Courvoisier}, {Foschini}, {Lebrun},
  {Lund}, {Parmar}, {Rodriguez}, {Tomsick} et~al.}]{2004ESASP.552..417W}
{Walter}, R., {Courvoisier}, T.~J.-L., {Foschini}, L., et~al. 2004.
\newblock In {Schoenfelder}, V., {Lichti}, G., \& {Winkler}, C., editors,
  \emph{5th INTEGRAL Workshop on the INTEGRAL Universe},  552 of \emph{ESA
  Special Publication}, 417--422

\bibitem[{{Walter} et~al.(2003){Walter}, {Rodriguez}, {Foschini}, {de Plaa},
  {Corbel}, {Courvoisier}, {den Hartog}, {Lebrun} et~al.}]{2003A&A...411L.427W}
{Walter}, R., {Rodriguez}, J., {Foschini}, L., et~al.
\newblock  2003, \aap,  411, L427--L432.
\newblock \doi{10.1051/0004-6361:20031369}

\bibitem[{{Watanabe} et~al.(2003){Watanabe}, {Sako}, {Ishida}, {Ishisaki},
  {Kahn}, {Kohmura}, {Morita}, {Nagase} et~al.}]{2003ApJ...597L..37W}
{Watanabe}, S., {Sako}, M., {Ishida}, M., et~al.
\newblock  2003, \apjl,  597, L37--L40.
\newblock \doi{10.1086/379735}

\bibitem[{{Watanabe} et~al.(2006){Watanabe}, {Sako}, {Ishida}, {Ishisaki},
  {Kahn}, {Kohmura}, {Nagase}, {Paerels} et~al.}]{2006ApJ...651..421W}
{Watanabe}, S., {Sako}, M., {Ishida}, M., et~al.
\newblock  2006, \apj,  651, 421--437.
\newblock \doi{10.1086/507458}

\bibitem[{{Wilms} et~al.(2000){Wilms}, {Allen}, and
  {McCray}}]{2000ApJ...542..914W}
{Wilms}, J., {Allen}, A., \& {McCray}, R.
\newblock  2000, \apj,  542, 914--924.
\newblock \doi{10.1086/317016}

\bibitem[{{Winkler} et~al.(2003){Winkler}, {Courvoisier}, {Di Cocco},
  {Gehrels}, {Gim{\'e}nez}, {Grebenev}, {Hermsen}, {Mas-Hesse}
  et~al.}]{2003A&A...411L...1W}
{Winkler}, C., {Courvoisier}, T.~J.-L., {Di Cocco}, G., et~al.
\newblock  2003, \aap,  411, L1--L6.
\newblock \doi{10.1051/0004-6361:20031288}

\bibitem[{{Yamaguchi} et~al.(2014){Yamaguchi}, {Eriksen}, {Badenes}, {Hughes},
  {Brickhouse}, {Foster}, {Patnaude}, {Petre} et~al.}]{2014ApJ...780..136Y}
{Yamaguchi}, H., {Eriksen}, K.~A., {Badenes}, C., et~al.
\newblock  2014, \apj,  780, 136.
\newblock \doi{10.1088/0004-637X/780/2/136}

\bibitem[{{Yaqoob} et~al.(2010){Yaqoob}, {Murphy}, {Miller}, and
  {Turner}}]{2010MNRAS.401..411Y}
{Yaqoob}, T., {Murphy}, K.~D., {Miller}, L., \& {Turner}, T.~J.
\newblock  2010, \mnras,  401, 411--417.
\newblock \doi{10.1111/j.1365-2966.2009.15657.x}

\end{thebibliography}
